\documentclass[fleqn,twoside]{article}
\usepackage[headings]{espcrc2}

\usepackage{graphicx}
\usepackage[figuresright]{rotating}

\newcommand{\AmS}{{\protect\the\textfont2
  A\kern-.1667em\lower.5ex\hbox{M}\kern-.125emS}}

\title{The chemical composition of the cosmic radiation around the ankle and the related
spectral indices}
\author{Antonio Codino \address[PG]{INFN and Dipartimento di Fisica dell'Universit\`a di
Perugia, Italy.} and Fran\c{c}ois Plouin\address[PG]{Former CNRS
researcher , Ecole polytechnique, LLR, F-91128, Palaiseau, France.}}

\begin{document}

\begin{abstract}

\par\parskip=1.truecm Some recent measurements of the chemical
composition of the cosmic radiation indicate that at the energy of
$3\times10^{18}$  eV,  around the ankle, light cosmic ions dominate
the spectrum as it occurs in the preknee energy region. \newline
Taking advantage of a recent theory of cosmic radiation which
provides a quantitative explanation of the knee, the second knee and
the ankle, the chemical composition of cosmic radiation is
explicitly calculated giving individual ion spectra and  ion
fractions  from 10$^{12}$ eV to $5\times10^{19}$ eV. The calculation
assumes two components of the cosmic radiation feeding the ion flux
at Earth: one originated in the disc volume and another one,  called
$\it {extradisc}$ $\it {component}$,  which from the disc boundaries
traverses the Galaxy reaching the solar system.   Data above
10$^{17}$ collected during half century of experimentation by Auger,
HiRes, Agasa, Akeno, Fly' s Eye, Yakutsk, Haverah Park and Volcano
Ranch experiments are reviewed, examined and compared with the
theoretical $<$$ln(A)$$>$.
\newline The comparison between computed and measured $<$$ln(A)$$>$
exhibits a good global accord up to $2\times10^{19}$ eV except with
the HiRes experiment and an excellent agreement in the range
$10^{15}$-$10^{17}$ eV with Kascade, Eas-top, Tunka and other
experiments. The accord requires a flux of the extradisc component
of $1.8\times10^{14}$ particles/($m^2$ sr s $eV^{1.5}$) at $10^{19}$
eV, twice that generated by disc sources.

\end{abstract}
\vspace{-0.5cm}

\maketitle


\section{Introduction}

The spectrum \footnote{ This paper is an english translation of an
INFN Report to be published in italian: $\it {Composizione }$ $\it
{Chimica}$ $\it {della}$ $\it {Radiazione}$ $\it {Cosmica}$ $\it
{intorno}$ $\it {alla}$ $\it {Caviglia}$ $\it {e}$ $\it {Indici}$
$\it {Spettrali}$ } of the cosmic radiation between the knee and the
ankle has been recently calculated according to the $\it {Theory}$
$\it {of}$ $\it {Constant}$ $\it {Spectral}$ $\it {Indices}$ [1-4].
The galactic mechanisms and the parameters generating the knee are
the same at the origin of the ankle: this circumstance constitutes a
major characteristic of the solution of the knee and ankle problem.
Another notable aspect of this solution is the presence of a
structure at $(5-7)\times10^{17}$ eV which is called the second
knee, just a direct consequence of the theory \cite{seconginoc}.
Figure 1 shows an example of excellent quantitative accord between
computed and measured cosmic-ray spectrum [4].


\begin{figure}[htb]
\vspace{-0.3cm}
\includegraphics [angle=0,width=8cm,height=8cm] {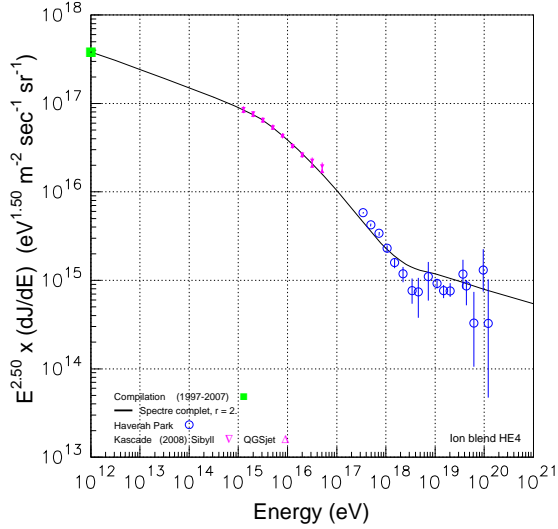}
\vspace{-2cm} \caption{ The differential
 cosmic-ray spectrum (black thick line) resulting from the theory normalized to
 $3.79 \times 10^{17}$ particles/($m^2$ sr s $eV^{1.5}$)
 at $10^{12}$ eV (green square). The data are from the Kascade $\cite{kaskaflussotot}$ and
  Haverah Park \cite{haverahtot}  experiments.}
\label{fig:largenenough} \vspace{-0.3cm}
\end{figure}


\begin{figure}[htb]
\vspace{-0.3cm}
\includegraphics [angle=0,width=8cm,height=8cm] {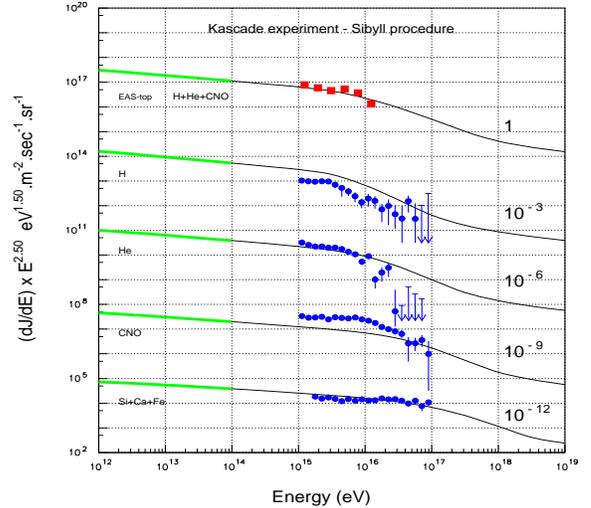}
\vspace{-1.5cm} \caption{Energy spectra of individual ions and group
of ions of Eas-top (red squares) $\cite{eastop1}$ and Kascade (blue
small circles) $\cite{kascaionspec}$ experiments compared with the
theory of constant spectral indices (black curve). Fluxes are
normalized at $10^{12}$ eV. Ion grouping and flux of the Eas-top
experiment shown in this figure  are examined elsewhere
$\cite{nota2flussidipro}$.} \label{fig:largenenough} \vspace{-0.4cm}
\end{figure}

 The $\it {Theory}$
$\it {of}$ $\it {Constant}$ $\it {Spectral}$ $\it {Indices}$
determines the energy spectra of individual ions, or group of ions,
in the interval $10^{11}$-$5\times10^{19}$ eV with only one
normalization point for the particle flux. As a consequence, once
the spectral indices and the abundances at a given energy are
assigned, the chemical composition of the cosmic radiation is known
at any energy.  Fig. 2 shows that individual ion spectra in the knee
energy region are also in good agreement with the theory. The
simplicity of the theory and  its anchorage to the empirical
parameters governing the cosmic-ion motion in the Galaxy are not
minor aspects besides the excellent quantitative agreement with ion
spectra exhibited in figures 1 and 2.
\par The
chemical composition expressed by $<$$ln(A)$$>$ has been also
calculated with all cosmic-ray sources placed in the disc
\cite{salina}, thus ignoring any extragalactic component. Figure 3
reports the results of the calculation.  The profiles of
$<$$ln(A)$$>$ measured by the Kaskade and Tunka experiments along
with that extracted with the measurements of $X_{max}$ of the Auger
Collaboration are shown in figure 3. There is an evident
disagreement above $10^{17}$ eV between theory and data. All other
experiments above $10^{17}$ eV have been purposely omitted from
figure 3 to render the disagreement between theory and data more
evident. According to the above quoted
 measurements the theoretical chemical composition,  as generated
by the galactic cosmic-ray sources above $4\times$$10^{17}$ eV,  is
incompatible with the data, primarily because it is too heavy, and
secondly, because  above $2\times$$10^{19}$ it has a flat trend,
contrary to the measured $<$$ln(A)$$>$,  which increases with
energy.


\begin{figure}[htb]
\vspace{-0.35cm}
\includegraphics [angle=0,width=8.5cm] {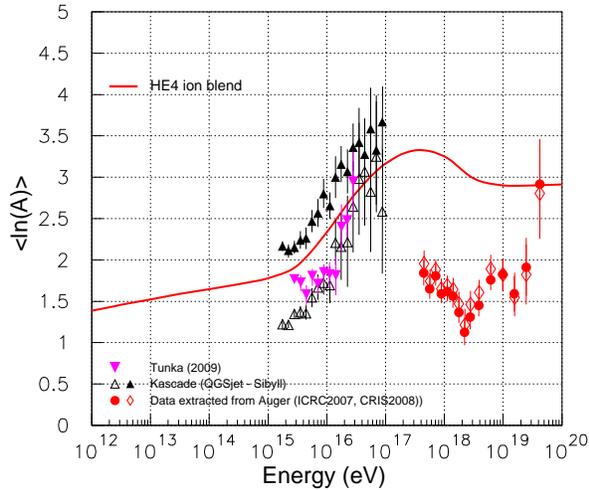}
\vspace{-1.5cm} \caption{Comparison of the $<$$ln(A)$$>$ derived
from the $\it {Theory}$ $\it {of}$ $\it {Constant}$ $\it {Indices}$
with the Tunka \cite{prosintunkaloga},  Kascade
\cite{kaskalogaorand} and Auger \cite{augerloga}  data. A
significant discrepancy between this calculation and the Auger data
appears above $4 \times 10^{17}$ eV. The theoretical evaluation of
$<$$ln(A)$$>$ (blue line) assumes that all cosmic-ray sources are
placed in the disc volume, with no extragalactic component.}
\label{fig:largenenough} \vspace{0.8cm}
\end{figure}

Notice that the Auger Collaboration disposes today (2009) of a
powerful and redundant  apparatus  and a number of recorded
atmospheric cascades superior to that of any other experiment.

The purpose of this work is to calculate the chemical composition of
the cosmic radiation at Earth,  not only assuming cosmic-ray sources
in the disc,  but also postulating a component that penetrates the
disc volume from the exterior reaching the Earth. This cosmic-ray
component is called here $\it {extradisc}$ being a particular
component of the more vast class of the extragalactic cosmic rays.

 Both the extradisc
cosmic rays and their flux will be denoted by the same symbol $\it
I_{ed}$ ($\it ed$ for extradisc). Let us anticipate here that $\it
{extradisc}$ $\it {component}$ reconciles the results of the theory
with the Auger data on $<$$ln(A)$$>$ shown in figure 3,  still
preserving a quantitative accord with the knees, the ankles and the
second knee.

\begin{table}[htb]
\begin{center}
\caption{ Parameters of the theory denoted Low Energy ($LE$) and
High Energy ($HE$) ion blends. Fluxes are multiplied by $E^{2.5}$
and expressed in units  of $(m^{-2} sr^{-1} s^{-1} eV^{1.5})$.}

\begin{tabular}{lrrrr}
\hline \vspace{-0.05cm}
Blend.        & LE              &          & HE              &         \\
\hline \vspace{0.2cm}
              & $10^{12}$ eV    &          & $10^{14}$ eV    &         \\
\hline \vspace{0.2cm}
              & \%              & $\gamma$ & \%              & $\gamma$ \\
\hline \vspace{0.2cm}
H             & 42.4            & 2.77     & 33.5            & 2.67     \\
He            & 26.5            & 2.64     & 27.3            & 2.64     \\
CNO           & 11.9            & 2.68     & 13.6            & 2.65     \\
Ne-S          &  9.2            & 2.67     & 10.9            & 2.63     \\
Ca(17-20)     &  1.2            & 2.67     &  2.8            & 2.63     \\
Fe(21-28)     &  8.7            & 2.59     & 12.0            & 2.62     \\
\hline
              & Flux            & $\gamma$ & Flux            & $\gamma$ \\
\hline
H             & $1.15\ 10^{17}$ & 2.77     & $3.93\ 10^{16}$ & 2.67     \\
He            & $7.19\ 10^{16}$ & 2.64     & $3.20\ 10^{16}$ & 2.64     \\
CNO           & $3.24\ 10^{16}$ & 2.68     & $1.60\ 10^{16}$ & 2.65     \\
Ne-S          & $2.50\ 10^{16}$ & 2.67     & $1.28\ 10^{16}$ & 2.63     \\
Ca(17-20)     & $3.14\ 10^{15}$ & 2.67     & $3.27\ 10^{15}$ & 2.63     \\
Fe(21-28)     & $2.36\ 10^{16}$ & 2.59     & $1.41\ 10^{16}$ & 2.62     \\
\hline
Total         & $2.71\ 10^{17}$ & 2.70     & $1.18\ 10^{17}$ & 2.64     \\
\hline
\end{tabular}
\end{center}
\vspace{-0.5 cm}
\end{table}

The structure of this paper preassumes the existence of the $\it
{Theory}$ $\it {of}$ $\it {Constant}$ $\it {Spectral}$ $\it
{Indices}$ [1-4] and the companion paper \cite{salina} where the
chemical composition of the cosmic radiation is evaluated under the
restricted assumption that all cosmic-ray sources feeding the
particle flux at Earth are placed in the disc volume. Section 2
presents a brief survey of the bases of the calculation and a flash
of the cultural background of the theory. Section 3 is devoted to a
survey on cosmic-ion energy spectra useful for the normalization of
the theory at $10^{12}$ eV. Section 4 summarizes the calculation of
the chemical composition obtained with the galactic sources only.
Section 5 reports a comparison between data and theory under the
assumption that all cosmic ray sources are placed in the disc and
not in its exterior. Section 6 reports many details of the
calculation of the ion abundances of the $\it {extradisc}$ $\it
{component}$ and the attenuation of its intensity while penetrating
the disc volume from its periphery up to the solar system.

The results of the calculation reported  in Section 6 are utilized
in Section 7,  where the $<$$ln(A)$$>$  for the disc and extradisc
components are given. The comparison of the predicted $<$$ln(A)$$>$
with the data for the two components $\it I_{ed}$ and $\it I_{d}$
(disc component)  are given in Sections 8 and 9. Section 10 marks
the disagreement on the chemical composition between the HiRes
experiment and others.  Conclusions are in the last Section 11.


\begin{figure}[htb]
\vspace{1.35cm}
\includegraphics [angle=0,width=8cm]{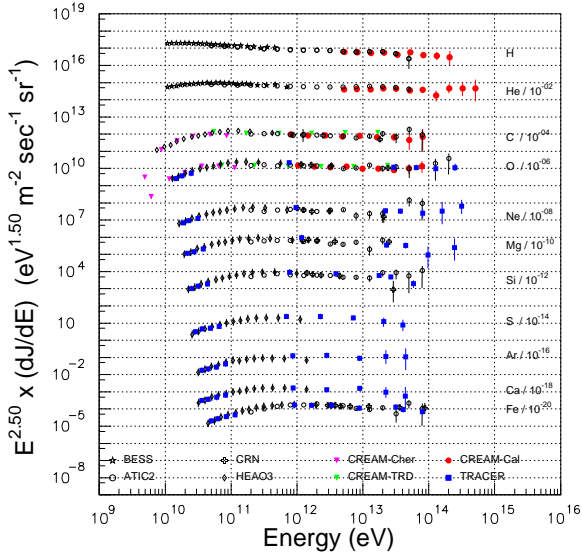}
\vspace{0.1cm} \caption{ Energy spectra of $11$ individual ions
measured by balloon and satellite detectors at energies below
$10^{15}$eV suggesting and founding the  postulate of the constant
spectral indices ( see ref. \cite{cod-merid} and references
therein).} \label{fig:largenenough} \vspace{0.1cm}
\end{figure}


\begin{figure}[htb]
\vspace{0.1cm}
\includegraphics [angle=0,width=8cm]{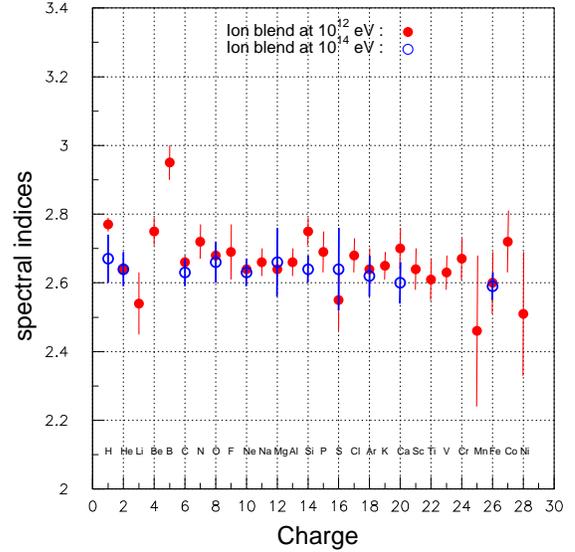}
\vspace{0.5cm} \caption{Spectral indices of cosmic ions from
Hydrogen to Nickel measured by balloon-borne and satellite detectors
at $10^{12}$ eV (filled red dots) and $10^{14}$ eV (open blue dots).
All indices extracted from the spectra around $10^{14}$ eV fall
between $2.6$ and $2.7$. } \label{fig:toosmall} \vspace{0.1cm}
\end{figure}


\begin{figure}[htb]
\vspace{-0.35cm}
\includegraphics [angle=0,width=8cm]{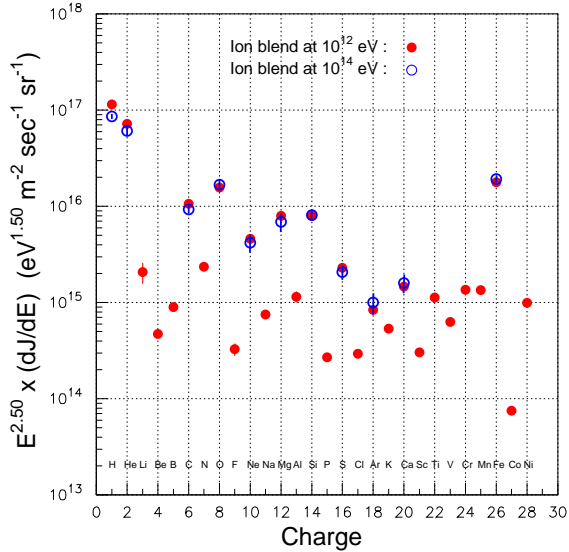}
\vspace{-1.5cm} \caption{Measured fluxes of individual ions of the
Wiebel-Sooth Compilation ($LE$ blend) (red dots) and $HE4$
Compilation (open blue dots).} \label{fig:largenenough}
\vspace{-0.7cm}
\end{figure}


\begin{figure}[htb]
\vspace{-0.3cm}
\includegraphics [angle=0,width=8cm] {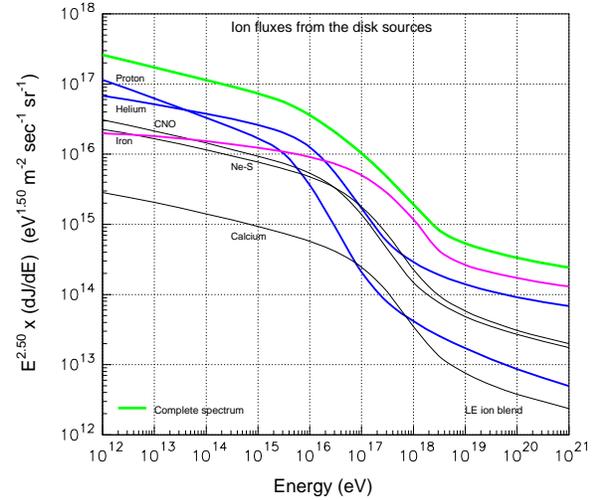}
\vspace{-1.5cm} \caption{  Theoretical spectra of 6 ions for the
$LE$ blend and the resulting $\it{complete}$ $\it{spectrum}$ (thick
green upper curve).} \label{fig:largenenough} \vspace{-0.4cm}
\end{figure}


\begin{figure}[htb]
\vspace{-0.3cm}
\includegraphics [angle=0,width=8cm]{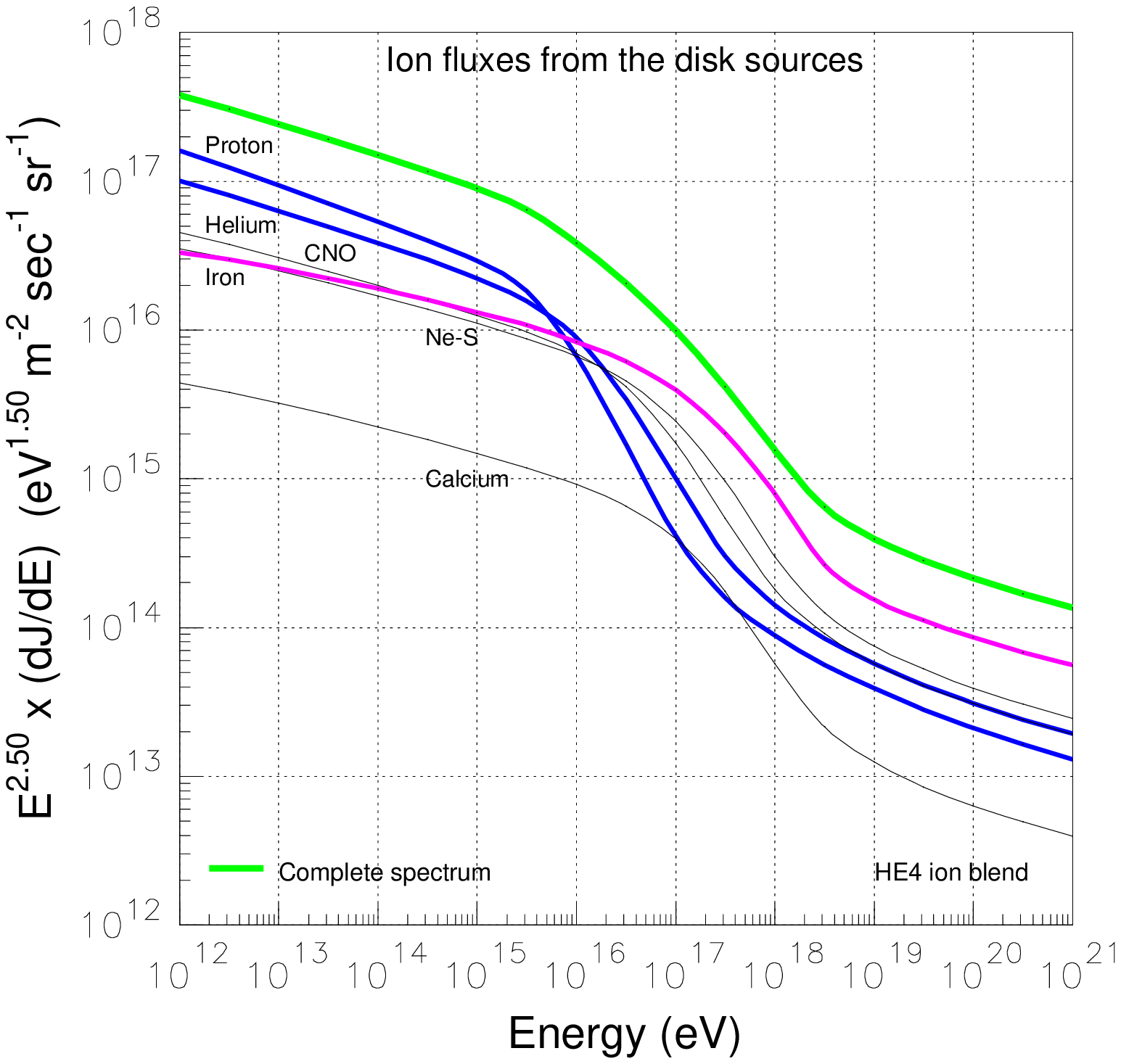}
\vspace{-2.1cm} \caption{ Theoretical spectra of 6 ions for the
$HE4$ ion blend and the resulting $\it{complete}$ $\it{spectrum}$
(thick upper green curve) normalized at $10^{12}$ eV with a flux of
$3.89\times10^{17}$ particles/$m^2$ sr s $eV^{1.5}$. }
\label{fig:largenenough} \vspace{-0.5cm}
\end{figure}


\section{Bases and foundation of the calculation }

A prerequisite inherent this calculation states  that cosmic rays
generated in the disc volume of the Milky Way might migrate into the
halo, and eventually,  into the intergalactic space [6]. The disc of
Milky Way is regarded as standard, typical disc volume. All these
cosmic rays escaped from the disc volume  populating the extradisc
space, constitute the $\it {extradisc}$ $\it {component}$. This
calculation admits that a fraction of the extradisc component
existing at the outskirts of the disc volume, by displacement
predominantly shaped by the the magnetic field and gas density, can
reach the solar cavity (the Earth): its intensity will denoted by
$\it I_{ed}$.

\begin{table}[htb]
\begin{center}
\caption{ Parameters of the theory at $10^{12}$ eV denoted universal
and $HE4$ ion blends. Fluxes are multiplied by $E^{2.5}$ and
expressed in units of $(m^{-2} sr^{-1} s^{-1} eV^{1.5})$.}

\vspace{1pc}
\begin{tabular}{lrrrr}
\hline
           &HE4           &         &Universal  &         \\
\hline
           & \%           &$\gamma$ & \%        &$\gamma$ \\
\hline
 H         & 42.4         & 2.74   & 37.2      & 2.645   \\
 He        & 26.5         & 2.72   & 26.4      & 2.640   \\
 CNO       & 11.9         & 2.69   & 13.9      & 2.650   \\
 Ne-S      &  9.2         & 2.68   & 10.1      & 2.650   \\
 Ca(17-20) &  1.2         & 2.67   & 01.5      & 2.650   \\
 Fe(21-28) &  8.7         & 2.66   & 10.8      & 2.650   \\
\hline
           &Flux          &         &Flux        &         \\
\hline
 H         &  1.61E+17    &         &  8.60E+16  &         \\
 He        &  1.01E+16    &         &  6.10E+16  &         \\
 CNO       &  4.54E+16    &         &  3.22E+16  &         \\
 Ne-S      &  3.50E+16    &         &  2.34E+16  &         \\
 Ca(17-20) &  0.44E+16    &         &  0.34E+16  &         \\
 Fe(21-28) &  3.31E+16    &         &  2.50E+16  &         \\
\hline
 Total     &  3.89E+17    &         &  2.31E+17  &         \\
\hline

\end{tabular}
\end{center}
\vspace{-0.5 cm}
\end{table}

In order to perform practical calculation it is necessary  to
specify some properties and features of the extradisc cosmic rays
present outside the disc:

(A) Cosmic rays emitted by the sources
placed in the disc and
  escaping from the disc volume undergo alterations of their
  abundances as released at the sources,  due to nuclear interactions
  with interstellar gas.
  It will be shown that this alteration depend both on the energy
  and the type of the ion.

(B) The spectral index of any ion released at the sources in the
disc volume is constant in the energy interval
$10^{11}$-$5\times10^{19}$ eV. The plausibility of this fundamental
hypothesis is rooted on the measurements of the spectral indices
collected in more than half a century by balloon and satellite
detectors (see fig. 4, 5 and 6). This hypothesis has been converted
in 2006 into a postulate called $\it {Postulate}$ $\it {of}$ $\it
{Constant}$ $\it {Spectral}$ $\it {Indices}$ \cite{origine}.

(C) Ion abundances as a function of energy of the extradisc
particles traversing the disc and arriving at Earth are evaluated
assuming the same parameters adopted for the disc component $I_d$.

  Though the present calculation ignores the mechanism that
  accelerate cosmic rays in the Milky Way,  two
  constraints have to be obeyed by the accelerator,  whatever it might
  be,
  in the range
  $10^{11}$-$5\times10^{19}$ eV : (1) spectral indices generated at the
  sources placed in the
  disc volume are independent of energy;
  (2) the cosmic-ray sources  are uniformly
  distributed in the disc volume at any energy \cite{Apjbrunetti-cod}.

  It has been demonstrated by full simulation of cosmic-ray motion
  that the displacement of the cosmic rays from the sources to any point in the Galaxy
  has a modest effect on the spectral indices released at the sources. For example,
  the propagation of helium in the disc entails an index modification
   of 0.04 in the energy band $10^{12}$-$10^{14}$ eV
  being at the sources 2.65. While for most calculations
this index modification  is a negligible effect,  for the evaluation
of the chemical
  composition around the ankle is of quite notable importance.

\par The interval of application of the $\it {Theory}$ $\it {of}$
 $\it {Constant}$ $\it {Spectral}$ $\it
{Indices}$  is $10^{11}$-$5\times10^{19}$ eV. The upper energy
extreme is determined by the realization that the characteristics of
the knee and the ankle of the $\it {complete}$ $\it {spectrum}$
(all-particle spectrum) are quantitatively explained by the theory.
The theory predicts the exact position where the ankle is observed,
at $(4-5)\times10^{18}$ eV and that of the knee as well, and many
other features. The quantitative characteristics of the ankle of the
$\it {complete}$ $\it {spectrum}$ derive from those of the ankles of
the individual ions. The proton ankle is  at the lowest energy while
the iron ankle is at the maximum energy, which turns out to be
$5\times10^{19}$ eV.

The theory distinguishes an ankle in the grammage versus energy and
an ankle in the cosmic-ray intensity versus energy. The knees are
intimately related by the theory to the ankles of any ions, and
since there exist an excellent quantitative agreement with numerous
experimental data,  the maximum energy involved in the knee-ankle
relationship is the Fe ankle energy. The Fe ankle energy in the
grammage is at $5\times10^{19}$ eV, and as a consequence, it
delimits the range of application of this theory. Above the energy
of $5\times10^{19}$ eV the results reported in this paper and the
companion one \cite{salina} have to be regarded as numerical
exercises. In fact, physical phenomena having empirical evidence
operate above $5\times10^{19}$ eV  but they are not incorporated in
the $\it {Theory}$ $\it {of}$ $\it {Constant}$ $\it {Spectral}$ $\it
{Indices}$  as it is presently formulated
\cite{{origine},{vulcano},{ginocchio},{traiettorie}}.

The results of the calculations depend on the geographic site of the
sources generating the extradisc component. For instance, if the
sources of the extradisc component would be the ensemble of the
spiral galaxies, due to the similarity of the Milky Way Galaxy, only
minor modifications of the calculations would be necessary to
determine $<$$ln(A)$$>$. If the sources of the extradisc  cosmic
rays would be placed in the local Superclaster of galaxies, the
present calculations are not sufficiently precise for a number of
reasons. If the extradisc component are just reentrant particles
escaped  from the disc of the Milky Way, as already suggested
\cite{vulcano}, the calculations are exact.

\par Total
kinetic
  energy of particles is used everywhere in this paper.

\begin{figure}[htb]
\vspace{-0.3cm}
\includegraphics [angle=0,width=8cm] {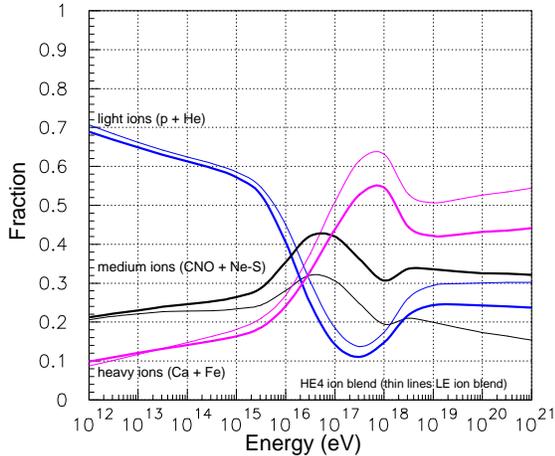}
\vspace{-2cm} \caption{ Relative amounts of light, intermediate and
heavy ions versus energy for the LE (thin curves) and HE (thick
curves) ion blends. Notice the minimum of the light component around
$3 \times 10^{17}$ eV in both ion blends, followed by a stable
increase of the light ion fraction up $4\times10^{18}$ eV.}
\label{fig:largenenough} \vspace{-0.6cm}
\end{figure}

\section{Measurements of the spectral indices and ion abundances }

\par Let us preliminarily notice that the interpolation of the
energy spectra of individual ions with a constant spectral index in
a restricted energy band it is an established tradition at energies
below $10^{15}$ eV. An index slightly depending  on energy would
probably conform better to some data,  in some energy intervals.
This refinement, however,  it is still uncommon, and unnecessary in
the present work, and consequently omitted here.

The fundamental hypothesis of the present calculation incorporated
in the $\it {Theory}$ $\it {of}$ $\it {Constant}$ $\it {Spectral}$
$\it {Indices}$ is that all ions at the sources have constant
spectral indices.

This simple hypothesis predominantly concurs in the prediction of
  the $\it {complete}$
 $\it {spectrum}$  leading  to the excellent
agreement with experimental data as shown, for example,  in figure
1.
\par Figure 4 reports the energy spectra of 11 different ions in the
interval $10^{10}$-$10^{15}$ eV.  The spectra form a grid of
parallel straight lines in the flux units shown. $\it {A}$ $\it
{posteriori}$  the regular forms of these spectra with no dips and
no spikes justify the postulate of $\it {Constant}$ $\it {Spectral}$
$\it {Indices}$. Though the evidence for constant indices is solid
and irrefutable below $5\times10^{16}$ eV there exist in $\it
{Nature}$ magnificent and universal phenomena that modify indices
released at the sources.  The ensemble of the processes referred to
as solar modulation occurring in all stars is an example of
violation of the $\it {Postulate}$ $\it {of}$ $\it {Constant}$ $\it
{Spectral}$ $\it {Indices}$ at low energy. Thus, local phenomena
evidently violates the postulate. In order to ascertain if
additional local phenomena in the Galaxy, besides solar modulation
and the knees,  in the interval $10^{11}$-$5\times10^{19}$ eV alter
the indices,  a careful examination of the experimental data is
mandatory.

A compilation of indices of a number of nuclides from Hydrogen to
Nickel is shown in figure 5. This compilation based on measurements
of the indices mainly between $10^{10}$-$10^{15}$ eV is referred to
as Wiebel-Sooth compilation \cite{wiebel-sooth}  and the
corresponding parameters incorporated in the theory are called $\it
{LE}$  $\it {ion}$ $\it {blend}$ (Low Energy). Another compilation
of spectral indices collecting measurements at higher energies
\cite{cod-merid} between $10^{12}$-$10^{15}$ eV is also shown in the
same figure 5 (blue open dots)and it is referred to as $\it {HE}$
$\it {ion}$ $\it {blend}$ (High Energy). The spectral indices shown
in figure 5 accumulate around a common value of 2.65 between
$10^{11}$-$10^{15}$ eV. It is also evident from the experimental
data on the indices shown in figure 5  (red dots) that higher the
atomic mass harder the index. This correlation between indices and
atomic mass is more pronounced for $LE$ than  $HE4$ ion blend.

Notice that the parameters of the Wiebel-Sooth Compilation
incorporated in the $LE$ ion blend based on the interval
$10\times$$Z$($GeV$) up to $10^{12}$ eV are three orders of
magnitudes from the knee energy region (e.g. $10^{15}$-$10^{18}$ eV)
while those of the $HE$ ion blend are closer. If the postulate of
the constant spectral indices is exactly described by a straight
line (in logarithmic scales of intensity and energy) either the $LE$
or $HE$ blend would suffice to correctly calculate the chemical
composition. In this circumstance the differences in the chemical
composition at any energy are only due to the ion abundances at the
normalization energy of the theory. If quite small violations of the
postulate take place, then, the differences originating from the two
ion blends become significant.

Recent experimental data between 40 and 400 $GeV$  indicates that
the proton index changes from 2.77 $\pm$ 0.02 to 2.67 $\pm$ 0.03 in
the interval $4\times 10^{11}$$4\times 10^{14}$ eV taking into
account the data of Atic2 \cite{atic2protoni} and Cream experiments
\cite{cream-merida}.

The hardening of the index seems to affect the helium spectrum as
well. In fact, the index of 2.70 $\pm$ 0.02 measured at low energy
below $400$ $GeV$ decreases to the value of 2.62 $\pm$ 0.05 above
$400$ $GeV$ up to the maximum energy explored to day of $4\times
10^{14}$ eV \cite{boyle}. The alteration of the H and He indices in
the preknee energy region would entail a change in the H/He flux
ratio which has been proved to be constant at low energy, $10^{13}$
eV \cite{goldenhedhe}. The H/He flux ratio versus energy has been
examined in another paper \cite{cod-merid} where it has been shown
that  the problem of the H/He flux ratio above $10^{13}$ eV, is, at
the present times,  inextricable. The recent data of the Cream
experiment \cite{cream2009lodz} do not resolve  the problem of the
H/He flux ratio below $10^{15}$ eV due to the large error bars of
the experiment at high energy, in the decade $10^{14}$-$10^{15}$ eV.

Figure 6 reports the ion fluxes for the 2 compilations defining the
parameters of the $LE$ and $HE$ blends. The fluxes are almost equal
for  the two blends as shown in fig. 6. The flux of the complete
spectrum  based on the Wiebel-Sooth compilation ($LE$ blend) is
$2.71\times 10^{17}$ particles/$m^2$ $s$ $sr$ $eV^{1.5}$ at
$10^{12}$ eV. The $HE$ blend at $10^{12}$ is normalized at the same
flux which at $10^{14}$ eV becomes $1.07\times 10^{17}$
particles/$m^2$ $s$ $sr$ $eV^{1.5}$. The complete spectrum has been
measured in the region $10^{12}$- $10^{15}$  giving a spectral index
of 2.74 and an extrapolated intensity at $10^{12}$ around $2\times
10^{17}$ particles/$m^2$ $s$ $sr$ $eV^{1.5}$ \cite{protonsatel}.

 In the
interval $10^{15}$-$10^{17}$ eV recent flux measurements  of the
complete spectrum by Kascade  give (Sibyll) $7.68\times 10^{16}$
 and (QGSjet) $7.48\times 10^{16}$ particles/$m^2$ $s$ $sr$
$eV^{1.5}$  at the reference (and arbitrary) energy of
$2\times10^{15}$ eV. Statistical errors of these measurements are
about 7 per cent. Note that the difference  in the $\it {complete}$
 $\it {spectra}$  resulting from QGSjet
and Sibyll  algorithms is negligible (see fig. 1). Differences in
flux measurements resulting from QGSjet and Sybill algorithms become
quite significant for individual ions amounting to factors  2-10 for
some ions in some energy intervals as pointed out elsewhere (see
Section 9 of ref. \cite{nota2flussidipro}).

The Kascade data suggest that the spectral indices of heavy ions are
constant in any preknee energy region. Empirically, they are
expected to be constant on the basis of balloon and satellite data
on proton and helium spectra collected below $10^{15}$ eV,  provided
that the H and He knees are regarded as experimentally observed,
around $2\times10^{15}$ eV and $6.7\times10^{15}$ eV, respectively.
If so, heavy ion spectra are expected to behave similarly.

The heavy ion spectra measured by the Kascade experiment do have
constant spectral indices close to 2.65
\cite{{kaskaionipes1},{kaskaionipes2}}. Any methods of analysis
(QGSjet, Sibyll or deconvolution method) yield indices of about 2.65
\footnote{ In the paper : $\it {The}$ $\it {Transition}$ $\it
{from}$ $\it {Tortuous}$ $\it {to}$ $\it {Rectilinear}$ $\it
{Cosmic}$ $\it {Ray}$ $\it {Trajectories}$ $\it {is}$ $\it {at}$
$\it {the}$ $\it {Origin}$ $\it {of}$ $\it {the}$ $\it {Knee}$ by A.
Codino and F. Plouin  \cite{traiettorie}, \quad the theoretical Fe
 spectrum is compared with that measured by the Kascade experiment,
which has an index of 2.64 $\pm$ 0.06. In this comparison heavy ions
include nuclei from Silicon to Nickel. Different groupings result in
similar indices but always in overabundant fluxes compared to the
theory (see figs. 16 and 17 of ref. \cite{seconginoc} ).} or harder
but not softer. When harder indices are observed a contamination
between nearby ions is suspected or demonstrated (see Section 7 of
ref. $\cite{nota2flussidipro}$ devoted to this argument). With the
appropriate ion groupings  of contaminated adjacent ion samples and
the correct guide from the theory [1-4], a constant common index for
heavy ion spectra in the fundamental energy region
$10^{15}$-$10^{17}$  emerges $\cite{nota2flussidipro}$.

Though the present calculation does not depend on any acceleration
mechanism, a theory of cosmic-ion acceleration in pulsar atmospheres
having constant spectral indices up to $5\times10^{19}$ eV has been
conceived in 1969 \cite{pizzella}.

The findings on the indices of heavy ion spectra of the Kascade
experiment consolidate the $\it {Postulate}$ $\it {of}$ $\it
{Constant}$ $\it {Spectral}$ $\it {Indices}$ from $10^{15}$ eV up to
about $10^{17}$ eV (unobserved Fe knee).  As far as the $\it
{Postulate}$ $\it {of}$ $\it {Constant}$  $\it {Indices}$ is at
stake,  the outcomes of the Kascade experiment on heavy ion spectra
in the interval $10^{15}$-$10^{17}$ eV have anticipated, probably by
more than half a century, those plausibly expected by future balloon
and satellite experiments, which identify individual ions with
superior discrimination power but with a very few nuclei.

\par  The data shown in figure 5 are included in the theory
with 6 indices and 6 fluxes at the normalization energy of $10^{12}$
eV.  The values of these parameters are given in Table 1.

In the same Table 1 are given the parameters of the $HE$ blend tuned
to the data at higher energy. In the following, in order to
appreciate  differences and similarities
 with previous calculations, the parameters of 2 additional ion
 blends referred to as universal
and $HE4$ blends are given in Table 2. In the universal ion blend
all indices are equal with a value of 2.65 except for helium set at
2.64 which takes into account the recent trend of the measurements
suggesting an index harder than that of other ions. The $HE4$ blend
is a variant of the $HE$ blend where the dominant nuclei, e.g.
protons and helium, H and He have the classical value of 2.74
\cite{protonsatel} and 2.72 \cite{jace2.72} believed for many years
to be standard, reliable measurements. The parameters of the $HE4$
blend are used to determine the chemical composition around the
ankle (see fig. 6). The $HE4$ blend is normalized at $10^{12}$ eV
with a flux of $3.79\times10^{17}$ particles/$m^2$ $s$ $sr$
$eV^{1.5}$.  This value is inspired from the Kascade flux
measurement which is $7.58\times10^{17}$ particles/$m^2$ $s$ $sr$
$eV^{1.5}$ at $2\times10^{15}$ eV  (mean value of QGsjet and Sibyll
algorithms) \cite{kaskaflussotot}. The $HE4$ flux normalization at
$10^{12}$ eV is larger than that of the $LE$ blend (Wiebel-Sooth
flux compilation) by a factor 1.4.

Because of the ion propagation in the disc volume the indices of the
original spectra released by the sources modify (see fig. 6 of ref.
3). It may be useful to subdivide the index of any ion spectrum
measured at the solar cavity in two parts: the propagation index and
the accelerator index at the sources. The propagation indices due to
ion displacement in the disc volume for the 6 ions H, He, N, Si, Ca
and Fe are, respectively, 0.060, 0.055, 0.045, 0.045, 0.040 and
0.040. Taking into account the accelerator indices  of the universal
blend (see Table 2)  the global index become: 2.585, 2.585, 2.605,
2.605, 2.610 and 2.610 for the same ion sequence just mentioned.


\section{Chemical composition of the cosmic radiation generated by
cosmic-ray sources placed in the disc}

 The energy spectra of 6 ions derived from the $\it {Theory}$ $\it {of}$
 $\it {Constant}$ $\it {Spectral}$ $\it
{Indices}$ with the $LE$ and $HE4$ blends are,
 respectively, in figure 7 and 8. The $\it {complete}$ $\it {spectrum}$
 which is the sum of the $\it {partial}$ $\it {spectra}$ of individual ions is also shown
 in the same figures 7 and 8 for both blends resulting in
 a flux of $8.2\times 10^{16}$
particles/$m^2$ $s$ $sr$ $eV^{1.5}$ for the $HE4$ ion blend at
$10^{15}$ eV. The ion abundances
 derived from the energy spectra
 reported in figures 7 and 8 are shown in figure 9 for the interval
$10^{12}$-$5\times 10^{19}$   eV. The light ion fraction
 (H+He)  decreases with a gentle slope going from 68
 per cent at $10^{12}$ eV
 to 55 per cent at $1.5\times 10^{15}$ eV ($HE4$ ion blend).
  Above this energy the light ion fraction descends abruptly reaching a
   minimum of 11 per cent at the energy of $4\times 10^{17}$ eV,
   quite evident in figure 9. A similar behaviour is exhibited by
   the $LE$  blend but with higher light ion fractions.

The slight increase in the (Ca+Fe) fraction  in the interval
$10^{12}$-$4\times 10^{15}$ eV for the $HE4$ blend is caused by the
tiny difference in the Ca and Fe slopes (2.66 for Fe and 2.650 for
Ca,  see Table 2) with respect to the light ion slopes (2.74 for H
and 2.72 for He). The strong increase of the (Ca+Fe) fraction
between $4\times 10^{15}$ eV and $7\times 10^{17}$ eV is due not
only to the difference in the indices(2.70 light ions against 2.60
heavy ions) but predominantly to the high fall of the light ion flux
(light ion knees) above the nominal knee of the complete spectrum at
$3\times 10^{15}$ eV. The fall is quite notable for protons and
helium in the interval $2\times 10^{15}$-$2\times 10^{17}$ eV but
insignificant for heavy ions (Ca+Fe) at least up to $4\times
10^{17}$ eV. The processes causing this abrupt fall of intensity
(knees) have been analyzed and described elsewhere \cite{ginocchio}.


\section{The  $<$$ln(A)$$>$   of  cosmic-ray disc component
 compared with the data}

\par  It is instructive to compare the theory with the experimental data
on $<$$ln(A)$$>$ under the restricted assumption that all cosmic-ray
sources are in the disc volume with no extragalactic component.


\begin{figure}[htb]
\vspace{-0.3cm}
\includegraphics [angle=0,width=8cm] {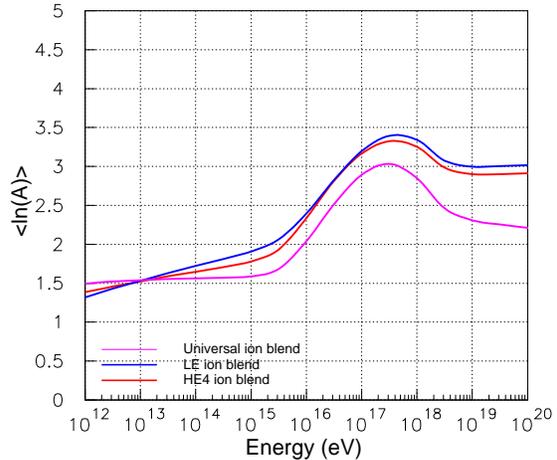}
\vspace{-2cm} \caption{Mean logarithmic mass of the cosmic radiation
resulting from the $\it {Theory}$ $\it {of}$ $\it {Constant}$ $\it
{Indices}$  in the energy interval $10^{11}$-$5\times 10^{19}$ eV
for three ion blends. In these 3 evaluations all cosmic-ray sources
are placed in the disc volume.} \label{fig:largenenough}
\vspace{-0.6cm}
\end{figure}


\begin{figure}[htb]
\vspace{-2.3cm}
\includegraphics [angle=0,width=8cm] {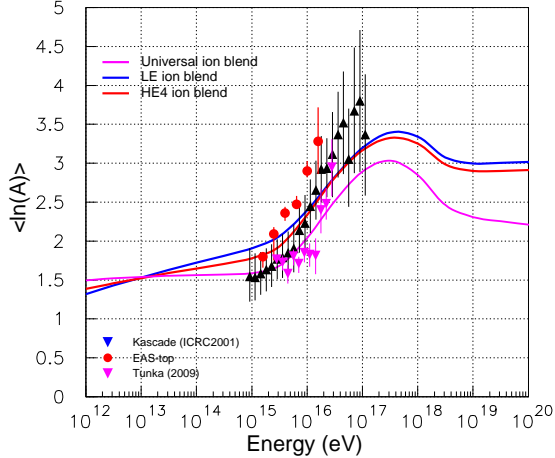}
\vspace{-0.6cm} \caption{Comparison of the theoretical $<$$ln(A)$$>$
for the $LE$ and $HE4$ blends with that measured by Kascade (black
triangles) \cite{datadeconvolution} using the deconvolution method
and that measured by Eas-top (red dots) \cite{eastoploga}. }
\label{fig:largenenough} \vspace{-0.6cm}
\end{figure}

 The ion spectra shown in figure 7 or 8  directly determine
 the chemical composition of the cosmic radiation
 in terms of $<$$ln(A)$$>$. The $<$$ln(A)$$>$ at energy
$E$ is the weighted
 average of the
 logarithmic atomic masses of the cosmic rays:


\begin{figure}[htb]
\vspace{-2.3cm}
\includegraphics [angle=0,width=8cm] {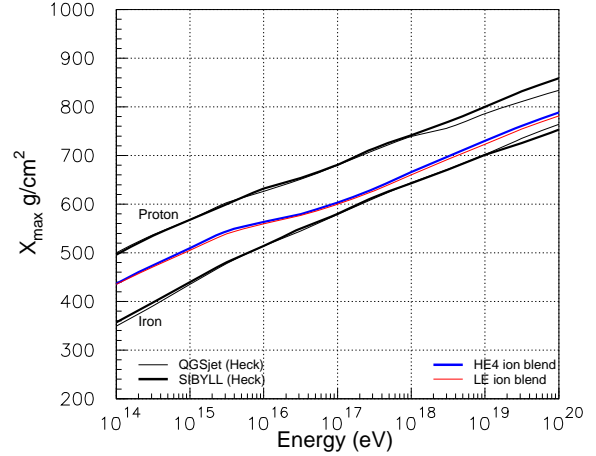}
\vspace{-0.6 cm} \caption{Atmospheric depths of shower maxima of
protons and Fe nuclei versus energy computed by  QGSjet and Sibyll
hadronic codes
 as evaluated by Heck \cite{heck}. }
\label{fig:largenenough} \vspace{-0.6cm}
\end{figure}


\begin{figure}[htb]
\vspace{-2.3cm}
\includegraphics [angle=0,width=8cm] {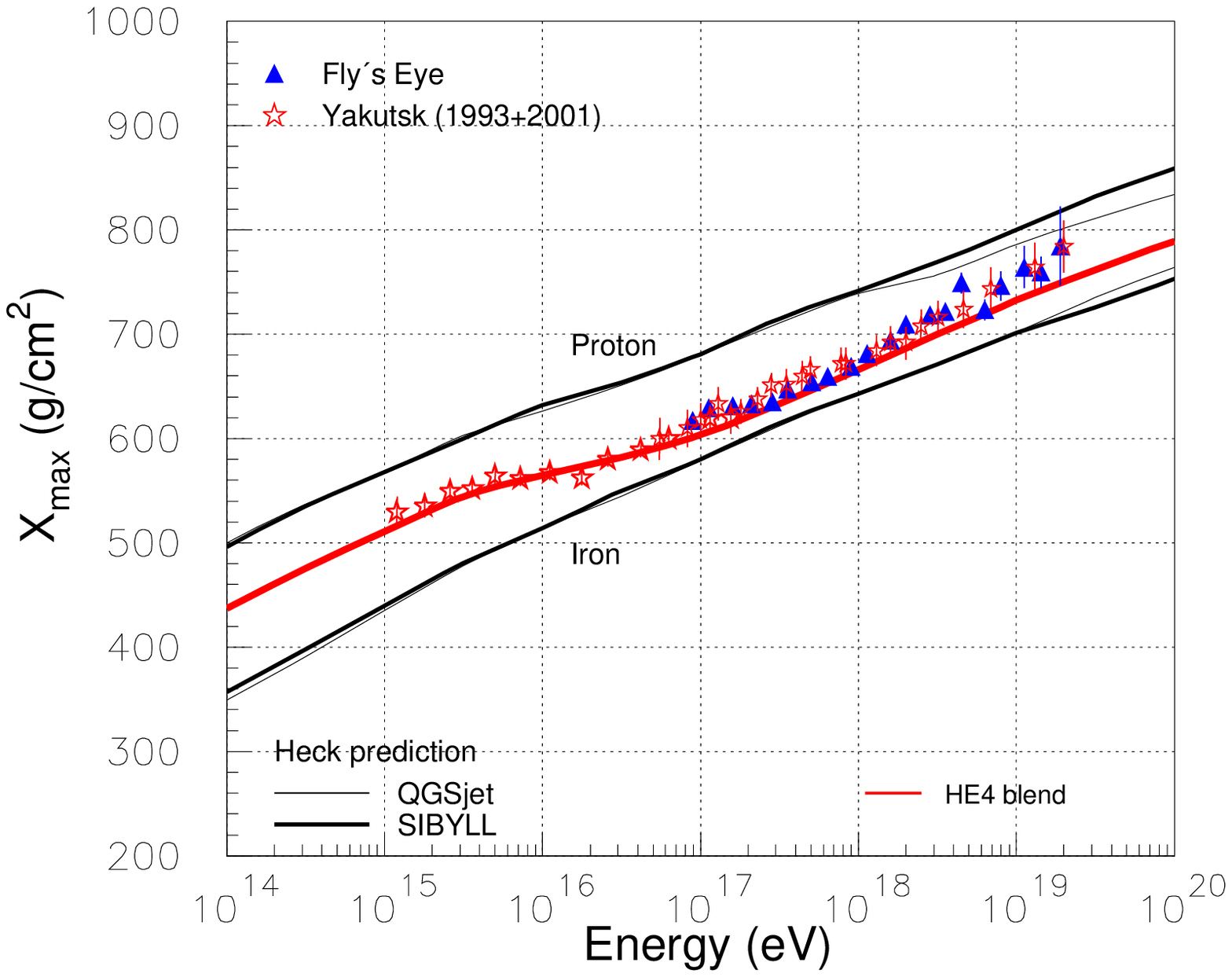}
\vspace{-0.6cm} \caption{Atmospheric depths of shower maxima
$X_{max}$ versus energy  measured by Fly's Eye (blue triangles)
\cite{occhidimosca} and Yakutsk (red stars) \cite{iachuscho}
experiments compared with the corresponding theoretical profile (red
line).  } \label{fig:largenenough} \vspace{-0.6cm}
\end{figure}

\parskip=0.15truecm
\quad \quad  $<$$ln(A)$$>$= $\sum$ $f_A$ $ln(A)$
\parskip=0.15truecm

where A is the atomic mass of
the nucleus and  $f_A$ its fraction in the cosmic-ray flux.


\begin{figure}[htb]
\vspace{-0.3cm}
\includegraphics [angle=0,width=8cm] {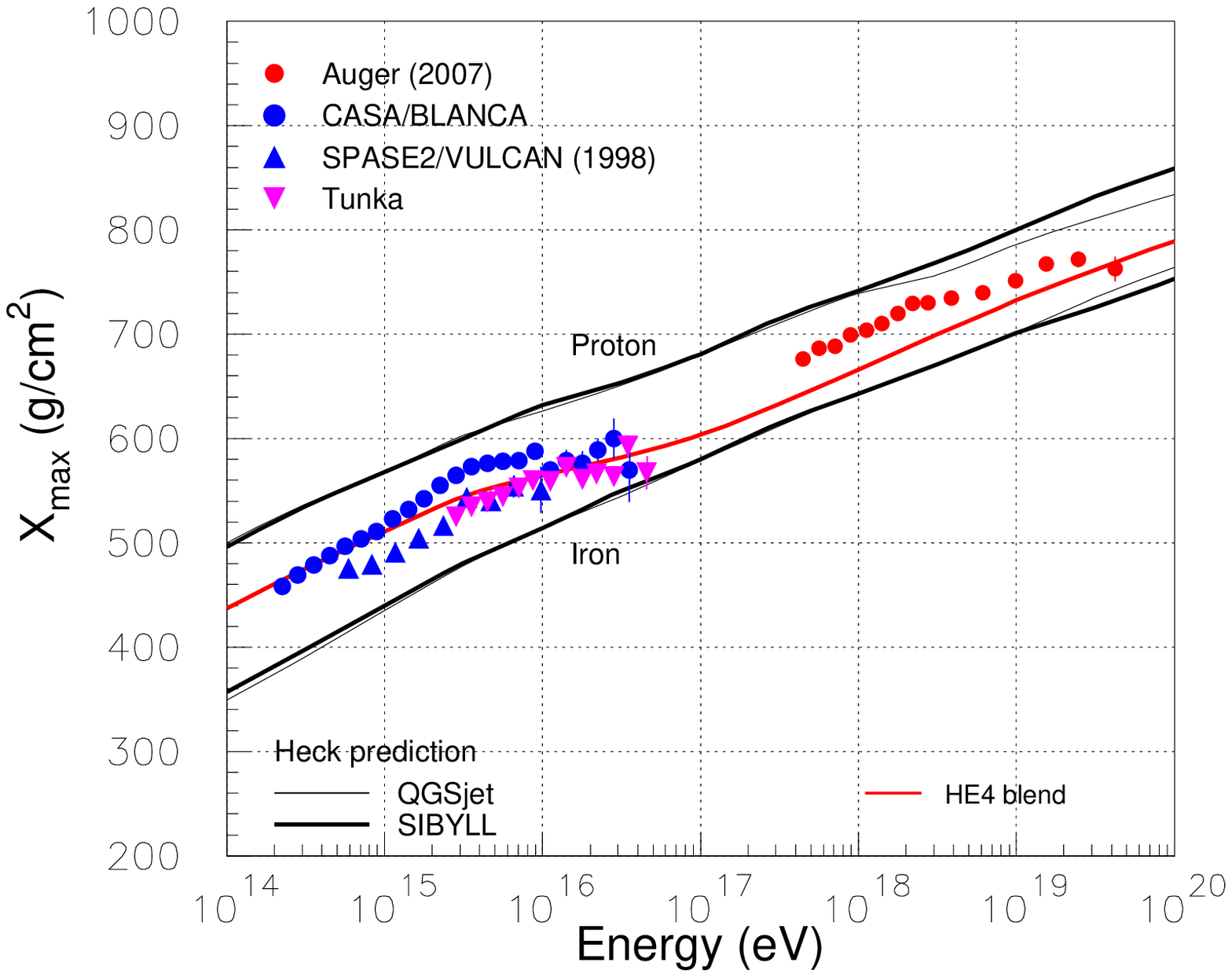}
\vspace{-1.6cm} \caption{Atmospheric depths of shower maxima of
protons and Fe nuclei versus energy measured by Auger, Space2/Vulcan
\cite{spacevulcan} and Casa-Blanca \cite{casablanca} experiments
compared with the  $<$$ln(A)$$>$ profile derived from the theory.}
\label{fig:largenenough} \vspace{-0.7cm}
\end{figure}


\begin{figure}[htb]
\vspace{-0.3cm}
\includegraphics [angle=0,width=8cm] {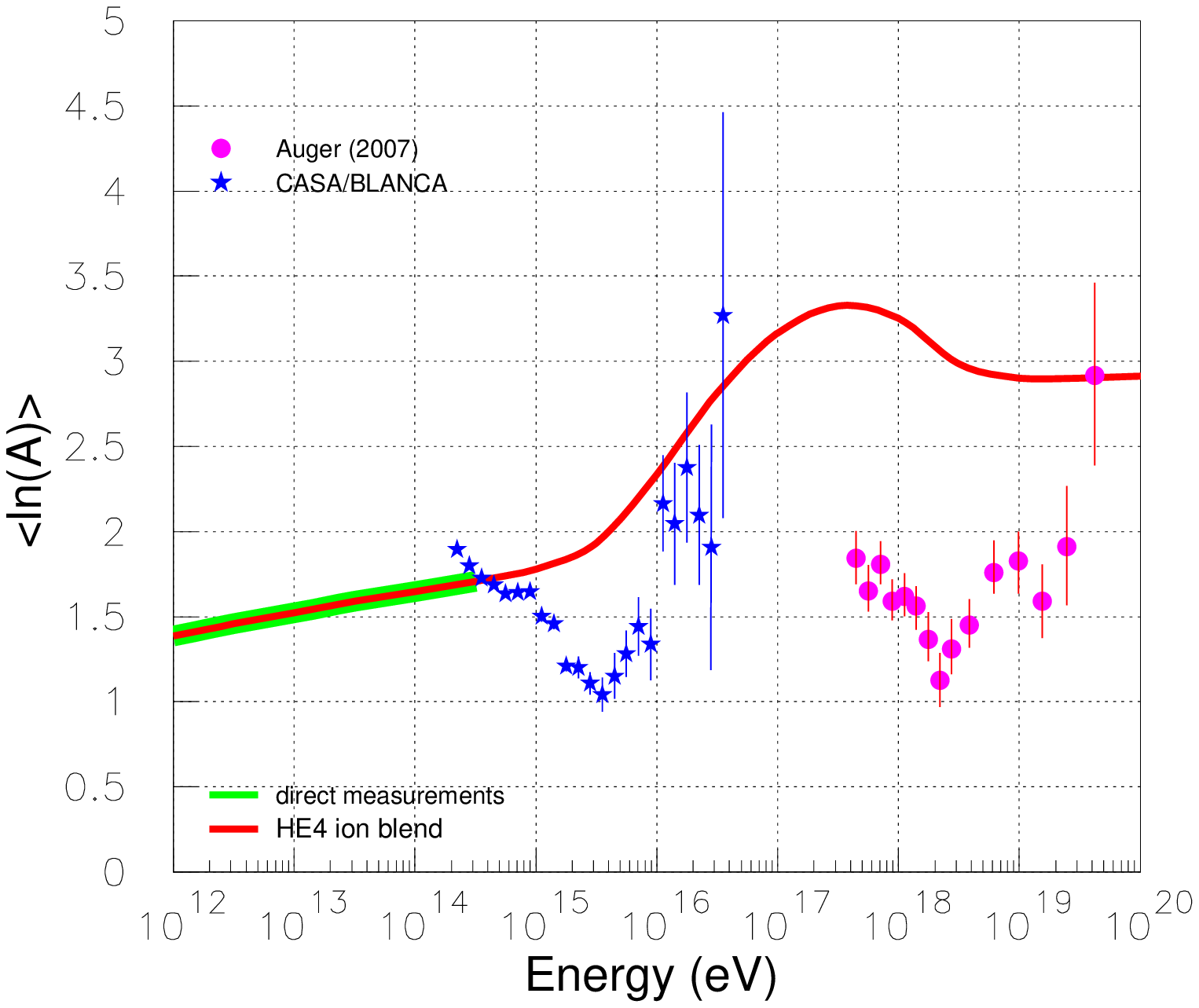}
\vspace{-1.6cm} \caption{Chemical composition in terms of
$<$$ln(A)$$>$  extracted by the $X_{max}$ measured by Casa-Blanca
\cite{casablanca} and Auger experiments.  The $<$$ln(A)$$>$ profiles
of each experiment curiously exhibit  prominent depressions at
$4\times10^{15}$ eV and $(2-3)\times10^{18}$ eV. The Casa-Blanca
depression is probably due to an instrumental effect as argued in
the text. } \label{fig:largenenough} \vspace{-0.7cm}
\end{figure}

In figure 10 are shown  3 profiles of $<$$ln(A)$$>$ resulting
 from 3 different ion blends: $LE$, $HE4$ and the universal blend.
The differences in the  $LE$ and $HE4$ blends  are quite negligible
in the entire interval $10^{12}$-$5\times 10^{19}$. There are
significant similarities but also some differences. The 3 profiles
are similar in the band $2\times 10^{15}$-$2\times 10^{17}$ eV, they
all  increase by about 2 units of $<$$ln(A)$$>$ and, in the range
$5\times 10^{17}$-$5\times 10^{18}$, they all descend by about one
unit. Differently, the $<$$ln(A)$$>$ profile of the universal blend
is flat from the initial energy at $10^{12}$ up to  $2\times
10^{15}$ eV, then it increases reaching a maximum of 3.05 at
$2.97\times10^{17}$ eV, and a more pronounced descent at high
energy, and then, above $5\times10^{18}$ eV, a modest decreasing
trend dominates. The form of the $<$$ln(A)$$>$ profile of the $HE4$
blend increases up to a maximum value of 3.33 at $3.94\times10^{17}$
eV, then it decreases rapidly up to $(4-5)\times10^{18}$ eV leveling
off at higher energies  around 2.8. As explained elsewhere (see fig.
4 of ref. \cite{origine}) the spectra of heavy ions (Ca and Fe)
attain the low asymptotic plateaux (see fig. 7 and 8) at higher
energies than the light ions. Low asymptotic plateaux reflect also
the quasi rectilinear propagation in the Milky Way.

Figures 11 and 3 show  data on $<$$ln(A)$$>$ of the Kascade
experiment which separates all nuclei in 5 groups along with the
corresponding theoretical curves ($LE$ and $HE4$). The Eas-top data
on $<$$ln(A)$$>$ are also shown in figure 11. Though there are
discrepancies in the ion spectra in figure 3 obtained by QGSjet and
Sibyll algorithms of the Kascade experiment \cite{nota2flussidipro},
the resulting $<$$ln(A)$$>$ profiles show the correct general trends
also observed by other experiments. By correct trend is meant an
average increase of  1.5 units of $<$$ln(A)$$>$ in the interval
$10^{15}$-$10^{17}$ eV from an initial value of 1.8 at  $10^{15}$ eV
dictated by the extrapolation of balloon and satellite data at lower
energies. The forms of the $<$$ln(A)$$>$ profiles in figure 11
obtained by the method of deconvolution \cite{haungsdeconvolution}
do not disagree with those of QGSjet and Sibyll algorithms shown in
figure 3, though the $<$$ln(A)$$>$ in some energy bands have notable
differences for the 3 methods. The agreement of the theoretical
$<$$ln(A)$$>$ profiles for the $LE$ and $HE4$ blends with the
Kascade data has been described in detail elsewhere \cite{salina}.

At energies larger than $10^{17}$ eV the determination of chemical
composition of the cosmic radiation becomes more involved because
ions are not resolved individually nor in restricted groups. In
order to determine the $<$$ln(A)$$>$ from the observables of the
atmospheric showers recorded by detectors,  accurate and detailed
models of the nucleus-air interactions are required. Traditionally,
the depth in $g$/$cm^2$ of the maximum of the atmospheric showers
(i.e. number of particles versus cascade axis)
 is denoted $X_{max}$ or equivalently
$\it elongation$.

A number of methods have been devised to measure $X_{max}$: (1) muon
number and muon density detected at ground; (2) the mean width of
the $X_{max}$ distribution denoted in short $\sigma({X_{max})}$; (3)
Cherenkov light generated in the cascades; (4) fluorescence light
generated in the cascades; (5) curvature of the cascade front
measured by time-of-flight techniques. Apparata exploiting Cherenkov
and fluorescence light can be equipped with additional detectors.

Figure 12 gives the $X_{max}$ versus energy evaluated by Heck
\cite{heck} for two models of nuclear interactions denoted in short
QGSjet and Sibyll. It results that the depth of the maximum of the
atmospheric showers induced by primary cosmic proton goes from 500
$g$/$cm^2$ at $10^{14}$ eV to 850 $g$/$cm^2$ at $10^{18}$ eV. In the
same energy interval the Fe elongation goes from 350 $g$/$cm^2$ at
$10^{14}$ eV to 750 $g$/$cm^2$ being the total thickness of the air
about
 1000 $g$/$cm^2$.

Above  $10^{14}$ eV  the gap between the proton $X_{max}$ profiles
of Sibyll and QGSjet regularly enlarges reaching 40 $g$/$cm^2$ at
$10^{20}$ eV (see fig. 12).

According to some simplifications the elongation is given by:

$$ X_{max} = D((ln E/E_0) -  ln(A)) + C  $$

 where  $X_{max}$ is the atmospheric depth in $g$/$cm^2$, for primary
 particles of energy $E$, $E_0$ is
 a reference energy, $<$$ln(A)$$>$ the chemical composition and $C$ and $D$ two
 appropriate functions. Let us note that in the present calculation,
 unlike others,
 $C$ and $D$ vary with energy. In the following, in the comparison
 with the experimental data, both $X_{max}$  and  $<$$ln(A)$$>$
 are maintained since the two variables are affected by small
 differences. The function $X^A_{max}$($E$,$E_0$) is the elongation
 for the nucleus $A$ with the energy
 $E$ normalized at the energy $E_0$.

The function $X^A_{max}$($E$,$E_0$)    obtained by
 cascade simulations depends on the hadronic models used to describe
 the interactions
 of cosmic nuclei with the air. $\quad$  Once the experimental value of
the elongation $X^{exp}_{max}$ has been determined,    $\quad$ the
value of $<$$ln(A)$$>$ is obtained in the superposition model by the
equation:

\par\parskip=0.6truecm
$<$$ln(A)$$>$= $ln (56)\times$
 ($X^{H}_{max}$-$X^{exp}_{max}$)$/$($X^{H}_{max}$ - $X^{Fe}_{max}$)
\par\parskip=0.6truecm

being  $X^{H}_{max}$  and $X^{Fe}_{max}$ the elongations simulated
by Monte Carlo for protons and Fe nuclei, respectively. Using the
above equation, the $<$$ln(A)$$>$ derived from the theory can be
converted into $X_{max}$. The result is shown in figure 12 with the
blue and red curves which are the average value of $X^{H}_{max}$ and
$X^{Fe}_{max}$ profiles obtained by QGSjet and Sibyll codes.

 The Yakutsk \cite{iachuscho} and
Fly's Eye \cite{occhidimosca} data on $<$$ln(A)$$>$ are shown in
figure 13 and they represent an example of accord between theory and
data. Note that at energies above $10^{18}$ eV the theoretical
$X_{max}$ profile has a substantial gap with the data. The accord
below $10^{18}$ eV and the discrepancy at high energy have been
discussed in the companion paper \cite{salina}.

Figure 14 reports  examples of disagreement.   The $X_{max}$
measured by Auger \cite{augerloga}  above $10^{17}$ eV exhibits a
large discrepancy between data and theory in the huge interval
$4\times10^{17}$- $10^{19}$ eV. Subsequent elaborations of the Auger
data samples  \cite{augerlodzloga}  mark the stability of the
$X_{max}$ except for the last data point which decreases by about 5
$g$/$cm^2$. The Casa-blanca and Space2/Vulcan data points in figure
14 have opposite deviations with respect to the theory, in different
energy intervals.

The measurements of the atmospheric depths of the cosmic radiation
made by Auger has been converted into $<$$ln(A)$$>$ and shown in
figure 15. According to these measurements cosmic rays around
$4\times10^{18}$ eV mainly consist of 37 per cent of heavy ions.
This figure comes from the partition of the cosmic nuclei in two
groups, light and heavy (see subsequent Section 8).

Besides the Auger experiment, a notable structure in the
$<$$ln(A)$$>$ profile, shown in figure 15,
 has been also
observed by the Casa-Blanca experiment in the interval $10^{14}$-
$4\times10^{16}$ eV. The depression of $<$$ln(A)$$>$ measured by
Casa-Blanca in the energy band $(3-9)\times10^{15}$ eV  would imply
the disappearance of heavy elements.  This depression corresponds to
the decreasing difference between the measured $X_{max}$ profile and
the theoretical $X^{H}_{max}$ in the interval
$2\times10^{14}$-$3\times10^{15}$ eV as shown in figure 14. As a
consequence, the depression of  the Casa-Blanca experiment shown in
figure 15 is not a spurious effect in the conversion of $X_{max}$
into $<$$ln(A)$$>$. This circumstance favours an instrumental effect
and not a physical cause for the origin of the depression. The
conclusion is corroborated by the measured by the Kascade
Collaboration with 3 methods (fig. 3 and 11) and others experiments
which, contrary to Casa-Blanca data, register the disappearance of
light elements.


\section{Ion abundances outside the disc volume}

 In order to determine the ion abundances in the space outside
     the disc volume it is necessary to calculate the probability of
     escaping from the disc,  $P_E$,  for any individual ions at all energies.
     At very high energy,  as ions propagate almost rectilinearly, this
     calculation is extremely simple.


\begin{figure}[htb]
\vspace{-0.3cm}
\includegraphics [angle=0,width=8cm] {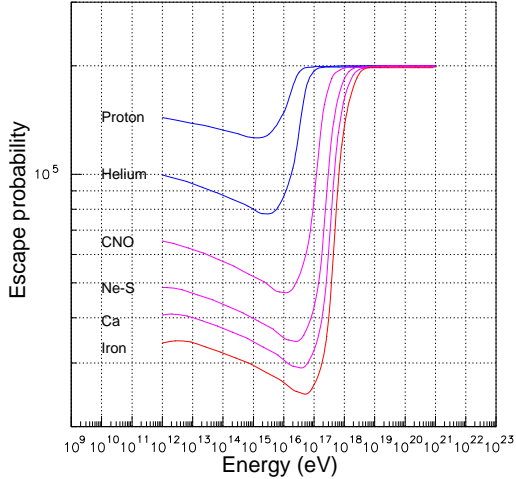}
\vspace{-1.6cm} \caption{Probability of escaping from the disc $P_E$
as a function of the energy for particles having cosmic-ray sources
distributed uniformly in the disc volume \cite{Apjbrunetti-cod}. The
probability $P_E$ is normalized to $2\times10^{5}$ particles. }
\label{fig:largenenough} \vspace{-0.7cm}
\end{figure}

Let $E^{d}_A$ be the energy
     above which  ions of mass $A$ propagate almost
     rectilinearly in the disc. In the rectilinear propagation
     regime ions encounter
 a minimum amount of matter as evaluated elsewhere \cite{ginocchio} .


\begin{figure}[htb]
\vspace{-0.3cm}
\includegraphics [angle=0,width=8cm] {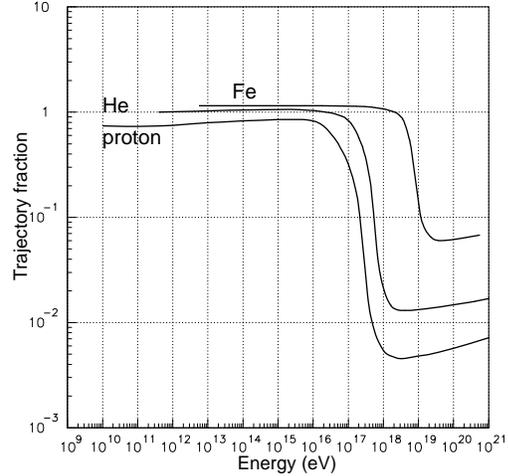}
\vspace{-1.6cm} \caption{Number of nuclear collisions versus energy
occurring in the disc volume for nuclei (H, He and Fe) having all
the initial points of the trajectories placed in the solar cavity.}
\label{fig:largenenough} \vspace{-0.7cm}
\end{figure}
The amount of matter experienced by ions while propagating in the
disc is referred to as $\it {grammage}$, $g$. It  is given by:
$g$=$m$$n$$L$,  where $m$ is the mean atomic mass in the disc, $n$
is the average number of atoms per $cm^3$ in the interstellar space
and $L$ the trajectory length. The grammage depends on nuclear cross
sections via trajectory length,  $L$. Since for $E$ $>$ $E^{d}_A$
the grammage is only 0.006 $g$/$cm^2$,  the number of nuclear
collisions taking place in the disc is negligible, and consequently,
ion abundances released by the cosmic-ray sources are almost
unmodified when they are observed in the solar cavity.

Unlike the rectilinear propagation regime  in the energy interval
where $E$ $<$ $E^{d}_A$,  between the knee and the ankle energy
region, e. g. $3\times10^{15}$-$5\times10^{18}$ eV, the galactic
magnetic field affects ion motion,  bending and inverting
efficiently ion directions. This effect of the magnetic field for
each ion makes the average grammage encountered by cosmic rays
dependent on particle momentum. For $E$ $<$ $E^{d}_A$, the
calculation of $P_E$ and the probability for cosmic ions of entering
(or re-entering) from the disc boundary to the solar cavity, $P_R$,
is more involved and difficult. In the following the functions $P_R$
are determined.

The number of particles escaping from the disc $N_F$ versus energy
for 6 ions or group of ions is given in figure 16. The quantity
$N_F$ is normalized at the same number ($2\times10^{5}$) of emitted
particles (becoming the probability $P_E$). All curves in figure 16
have a minimum at a particular energy $E^{min}_e$ ($e$ is for
escape). The characteristic energies where the minima of the 6 ions
H, He, N, Si, Ca and Fe occur are, respectively : $1.1\times10^{15}$
eV, $2.6\times10^{15}$ eV, $1.2\times10^{16}$ eV, $2.4\times10^{16}$
eV, $3.9\times10^{16}$ eV, and $4.5\times10^{16}$ eV.
 The
decreasing segment of $N_F$ from $10^{12}$ eV up to $E^{min}_e$ is
due to the increasing nuclear cross sections with energy while the
rising segment above $E^{min}_e$ is predominantly caused by the
vanishing efficiency of the galactic magnetic field with rising
energy to retain particles for long times.

Escaped particles populating the extradisc space may re-enter the
disc in the scheme delineated in Section 2, in a variety of
circumstances. Let be $P_R$ the mean probability for ions of energy
$E$ of reaching the Earth from the disc boundary. A cosmic-ray
trajectory consists of an initial point (the source) and a
termination point, which is the location where the nuclear collision
occurs within the disc volume. The calculation of $P_R$ exploits the
approximate symmetry of the cosmic-ray trajectories contained in the
disc volume between the initial point $X_I$ and the final point
$X_F$ of the trajectory. Samples of $2\times10^{5}$ particle are
injected from the solar system position and the trajectories
reconstructed in the disc volume. Let $N_I$ be the number of ions
stopping in the disc volume by nuclear collisions and $N_T$ the
total number of trajectories (or injected particles). The function
$N_I$ is of fundamental importance in the present calculation since
it incorporates the effect of the magnetic field, the dimension of
the disc, the variation of the nuclear cross sections with the
energy and the position of the solar system in the Galaxy. The
calculation is finally accomplished by taking advantage of the
inversion of the particle trajectories, the interchange of $X_F$
with $X_I$, as explained in detail elsewhere \cite{bacinicodplouin})


\begin{figure}[htb]
\vspace{-0.3cm}
\includegraphics [angle=0,width=8cm] {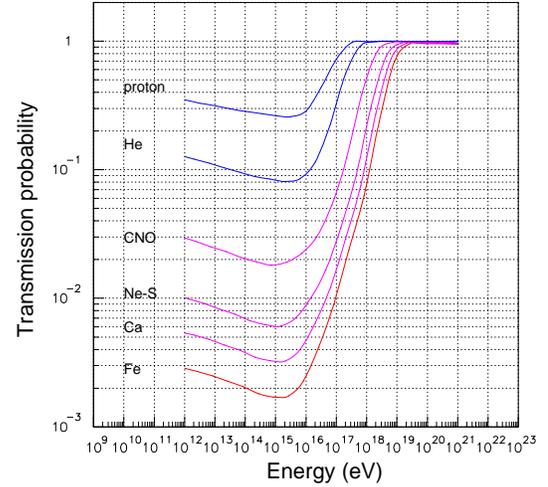}
\vspace{-1.6cm} \caption{Probability of reaching the solar system
for cosmic nuclei having  all the initial points of the trajectories
placed at the disc boundary.} \label{fig:largenenough}
\vspace{-0.7cm}
\end{figure}

Figure 17 reports the functions $N_I$ for 3 representative ions H,
He and Fe taken as reference examples. The $N_I$ profile of any ion
has a high plateau,  a quite small rise controlled by nuclear cross
sections, a rapid and high descend, a minimum, and finally,  a
modest increase up to the highest energies. For example, the number
of He interactions in the disc,  in the interval
$5\times10^{11}$-$10^{15}$ eV,  increases by 3.5 per cent. At
$10^{12}$ eV the number of  He particles interacting in the disc
volume is 173000 out of 200000 injected from the Earth (hence,
$N_T$=$2\times10^{5}$  and $N_I$=$1.73\times10^{5}$). At
$5\times10^{18}$ eV where the minimum of $N_I$ for He nuclei occurs,
the number of He interactions reduces to 2000 (hence $N_I$=2000).
Therefore, the probability for He particles of reaching the solar
cavity, having the initial points of the trajectories at the disc
boundary,  is : $P_R$ = $1$ - ($N_I$/$N_T$). With the figures given
above,  the probability $P_R$ is 0.11 at $10^{12}$ eV and 0.99 at
$10^{18}$ eV. By this procedure the 6 probability functions versus
energy $P_R$ are obtained and shown in figure 18. The number of
nuclear collisions taking place in the disc volume shown in figure
17 (see figure 6 of ref. \cite{origine}) is converted in terms of
probability $P_R$ in figure 18. The normalization of the functions
in figure 17 is explained in detail on the investigation on the
origin of the knee (Section 4 of ref. \cite{ginocchio}). Thus, the
functions in figure 17 are closer to the simulation of the
cosmic-ray trajectories and they facilitate the comprehension of the
processes causing the ion knees \cite{ginocchio}, while those in
figure 18 more specifically relate to the present study.


\begin{figure}[htb]
\vspace{-0.3cm}
\includegraphics [angle=0,width=8cm] {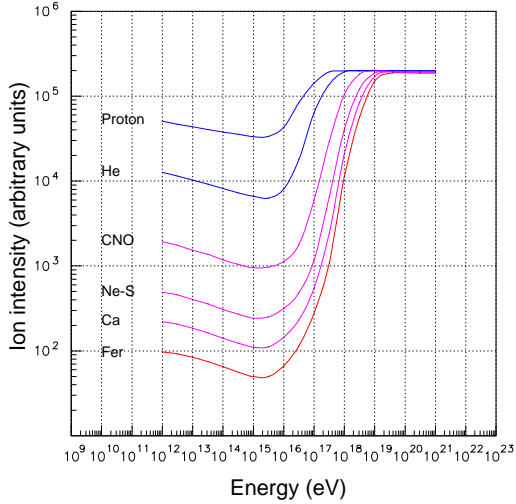}
\vspace{-1.6cm} \caption{Particle fraction, f,  versus energy of the
extradisc component reaching the solar cavity from the disc
boundaries (Halo of the Milky Way). The quantity $f$ is defined as
$f$=$N_e$$T$ where $N_e$ is the number of particles escaped from the
disc and $T$ the transmission probability (see fig. 18) from the
disc boundaries to the solar cavity at a specified energy. }
\label{fig:largenenough} \vspace{-0.7cm}
\end{figure}


\begin{figure}[htb]
\vspace{-0.3cm}
\includegraphics [angle=0,width=8cm] {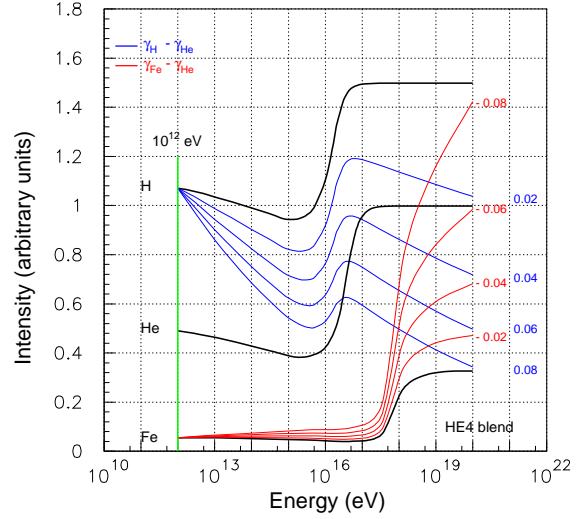}
\vspace{-1.6cm} \caption{Small differences of $\pm 0.1$ $\%$ in the
spectral indices at  $10^{12}$ eV will reverberate  large
alterations of the ion abundances  around the ankle energy region.
The grid of red and blue curves, specifying H and Fe ion abundances
versus energy, result from the change, by arbitrary steps of 0.02 of
the H and Fe indices,  with respect to the He index taken as
reference value (see text for details).} \label{fig:largenenough}
\vspace{-0.7cm}
\end{figure}


\begin{figure}[htb]
\vspace{-0.3cm}
\includegraphics [angle=0,width=8cm] {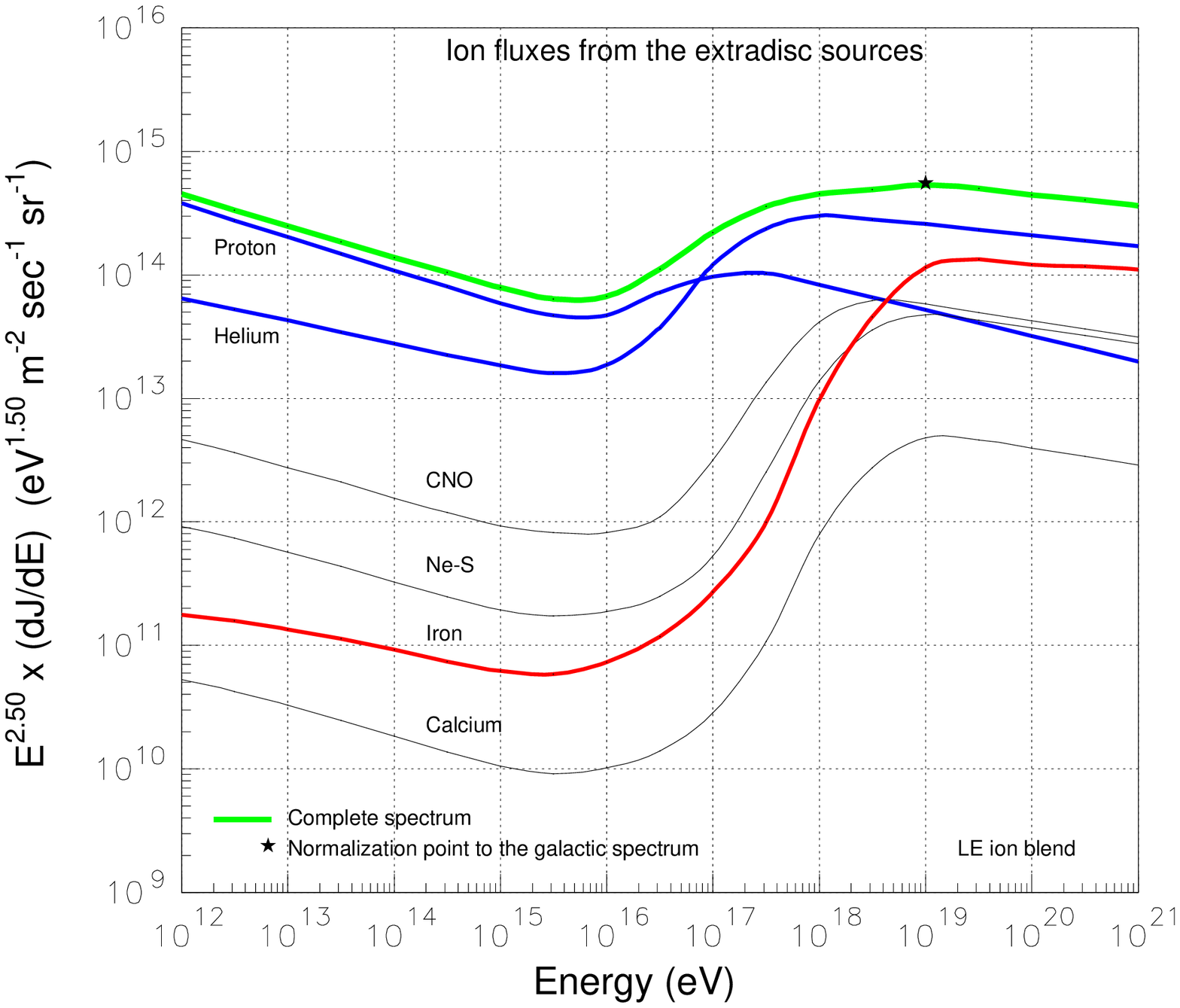}
\vspace{-1.6cm} \caption{Energy spectra of the extradisc ions
reaching the solar cavity emitted by sources characterized by the
$LE$ blend. The black star defines the flux normalization.}
\label{fig:largenenough} \vspace{-0.7cm}
\end{figure}


\begin{figure}[htb]
\vspace{-0.3cm}
\includegraphics [angle=0,width=8cm] {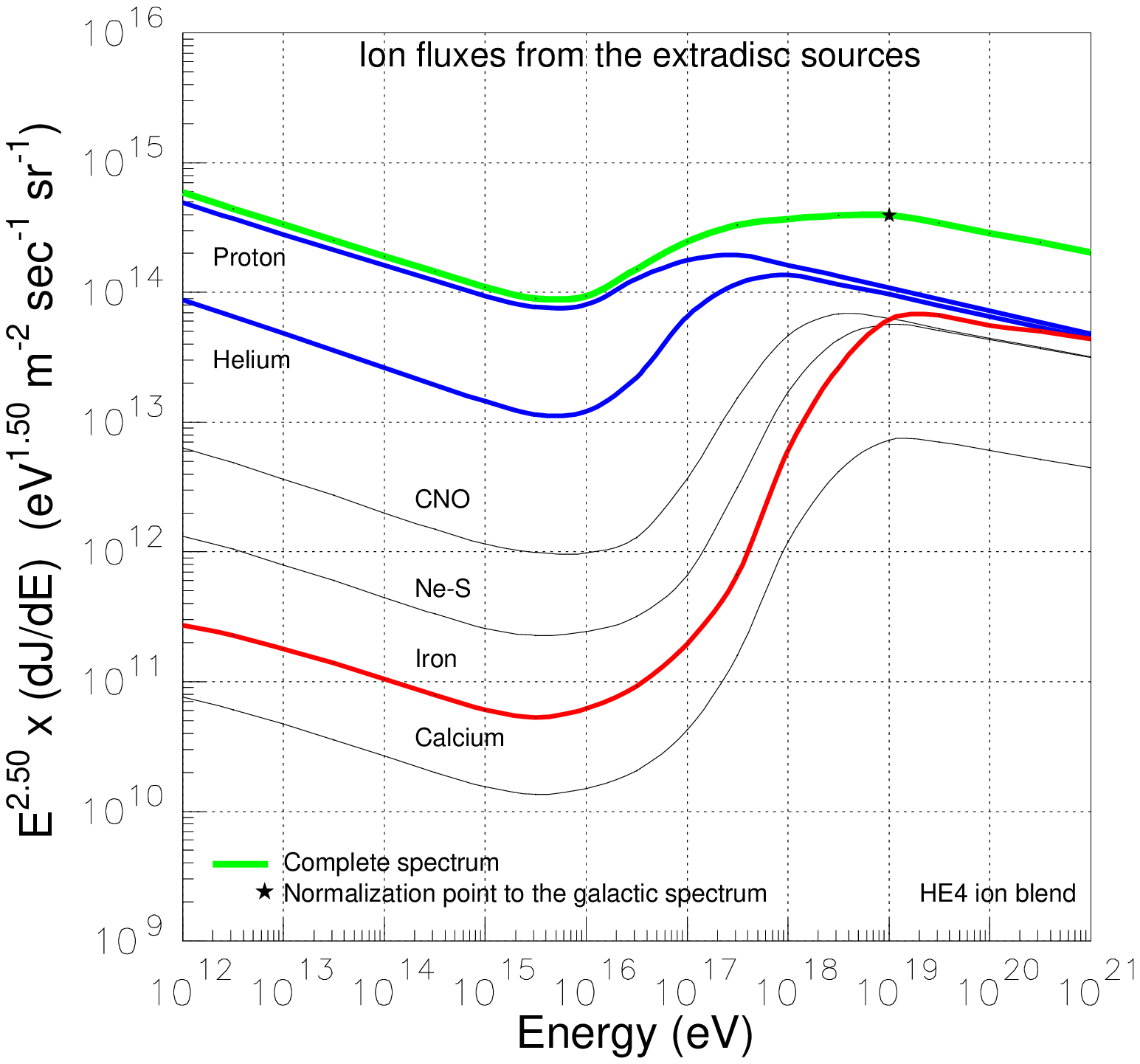}
\vspace{-1.6cm} \caption{Energy spectra of the extradisc ions
reaching the solar cavity emitted by sources characterized by the
$HE4$ ion blend. The black star signals the flux normalization.}
\label{fig:largenenough} \vspace{-0.7cm}
\end{figure}


\begin{figure}[htb]
\vspace{-0.3cm}
\includegraphics [angle=0,width=8cm] {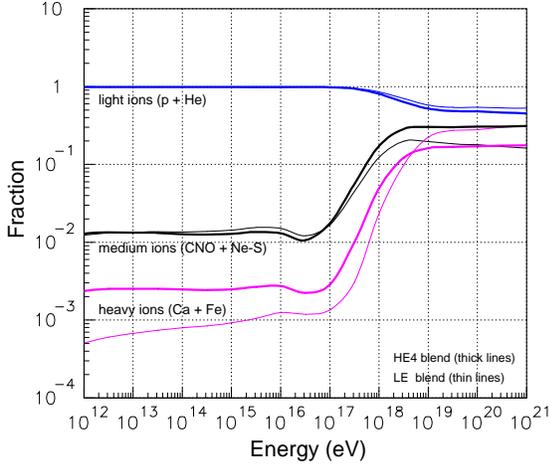}
\vspace{-1.6cm} \caption{ Relative amounts of light, intermediate
and heavy ions versus energy for the $LE$ (thin curves) and $HE4$
(thick curves) ion blends generated by the extradisc component.}
\label{fig:largenenough} \vspace{-0.7cm}
\end{figure}

\begin{figure}[htb]
\vspace{-0.3cm}
\includegraphics [angle=0,width=8cm] {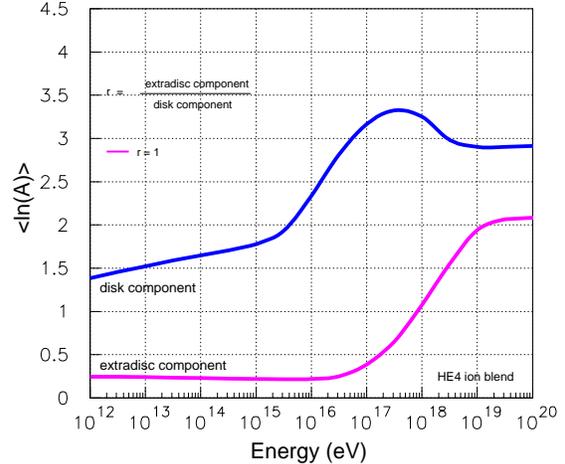}
\vspace{-1.6cm} \caption{Chemical composition in terms of
$<$$ln(A)$$>$ versus energy of the disc (blue curve) and extradisc
component (pink curve) for the $HE4$ ion blend derived from the
$\it {Theory}$ $\it {of}$ $\it {Constant}$ $\it {Spectral}$ $\it
{Indices}$. } \label{fig:largenenough} \vspace{-0.7cm}
\end{figure}


\begin{figure}[htb]
\vspace{-0.3cm}
\includegraphics [angle=0,width=8cm] {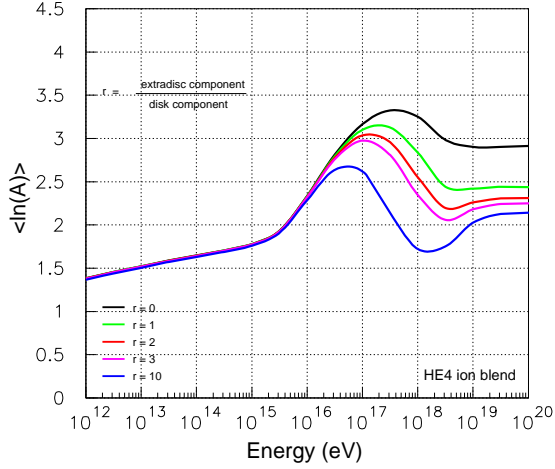}
\vspace{-1.6cm} \caption{ Chemical composition in terms of
$<$$ln(A)$$>$ versus energy for various values of the
extradisc-to-disc flux ratios, $r$,  for the $HE4$ ion blend.}
  \label{fig:largenenough}
\vspace{-0.7cm}
 \end{figure}


\begin{figure}[htb]
\vspace{-0.3cm}
\includegraphics [angle=0,width=8cm] {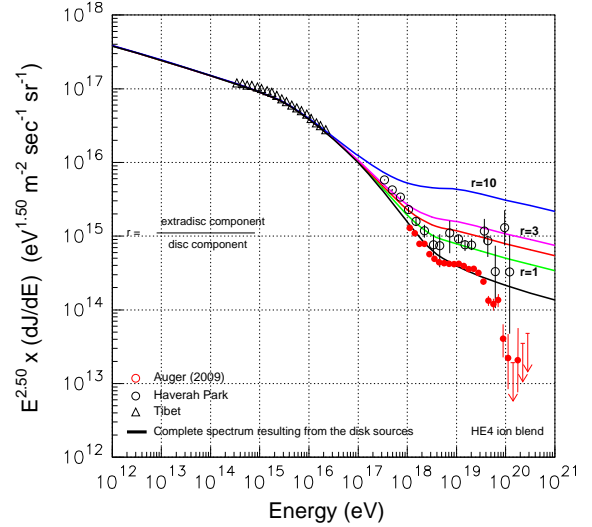}
\vspace{-1.6cm} \caption{ Energy spectra of the cosmic radiation
with different values of the extradisc-to-disc flux ratio, $r$,
derived from the $\it {Theory \quad of \quad Constant \quad Indices
}$ framed between the Haverah Park and Auger flux measurements.}
\label{fig:largenenough} \vspace{-0.7cm}
\end{figure}


\begin{figure}[htb]
\vspace{-0.3cm}
\includegraphics [angle=0,width=8cm] {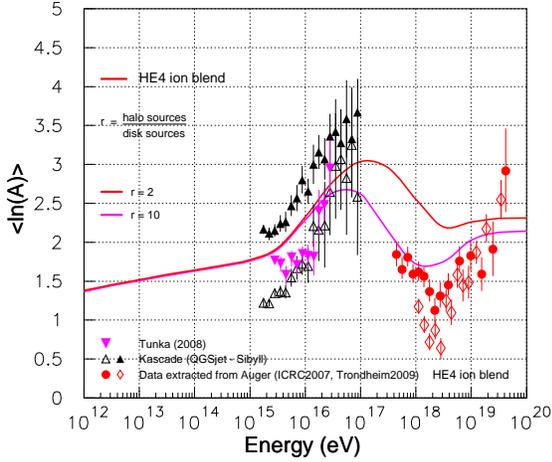}
\vspace{-1.6cm} \caption{ Comparison of the theoretical profiles of
 $<$$ln(A)$$>$ (red and pink curves) with the empirical one
 extracted from the $X_{max}$
measured by the Auger experiment (red diamonds and dots). The
extradisc-to-disc flux ratio at $10^{19}$ is 2 (blue line) and 10
(pink line).}
  \label{fig:largenenough}
\vspace{-0.7cm}
 \end{figure}


\begin{figure}[htb]
\vspace{-0.3cm}
\includegraphics [angle=0,width=8cm] {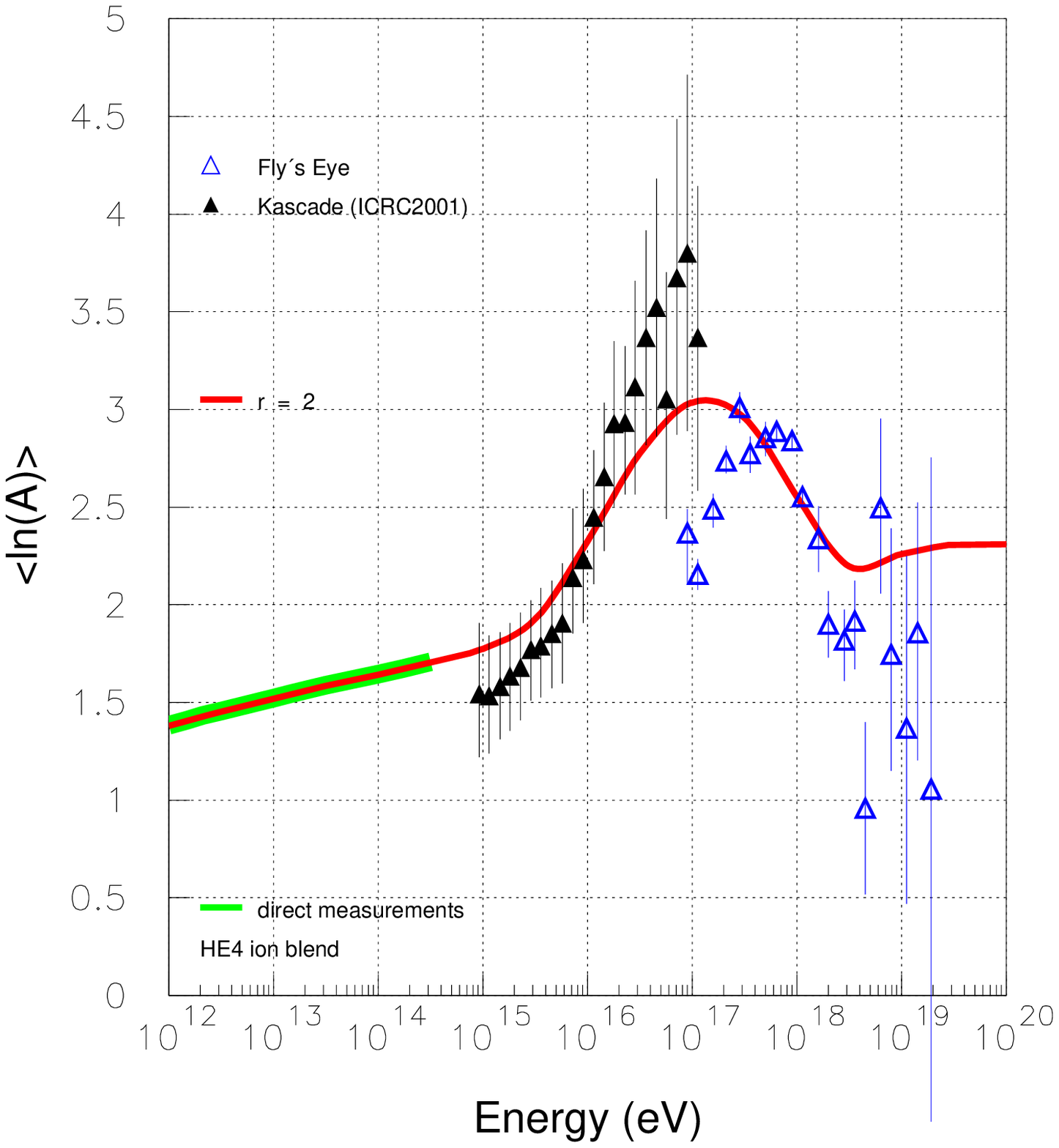}
\vspace{-1.6cm} \caption{ Comparison of the theoretical profile of
 $<$$ln(A)$$>$ (red curve) with the empirical one extracted from the data
 of Fly's Eye  experiment (blue triangles).}
  \label{fig:largenenough}
\vspace{-0.7cm}
 \end{figure}


\begin{figure}[htb]
\vspace{-0.3cm}
\includegraphics [angle=0,width=8cm] {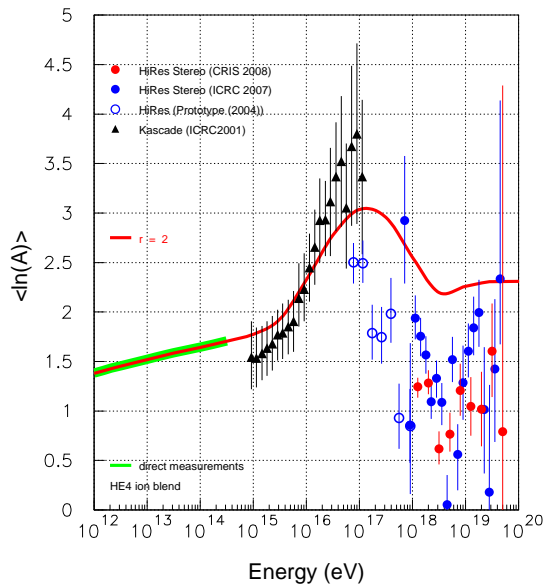}
\vspace{-1.6cm} \caption{ Comparison of the theoretical profile of
 $<$$ln(A)$$>$ (red curve) with the empirical one extracted from the data
 of HiRes experiment before (blue data points)and after
 (red data points) acceptance corrections.}
  \label{fig:largenenough}
\vspace{-0.7cm}
 \end{figure}


\begin{figure}[htb]
\vspace{-0.3cm}
\includegraphics [angle=0,width=8cm] {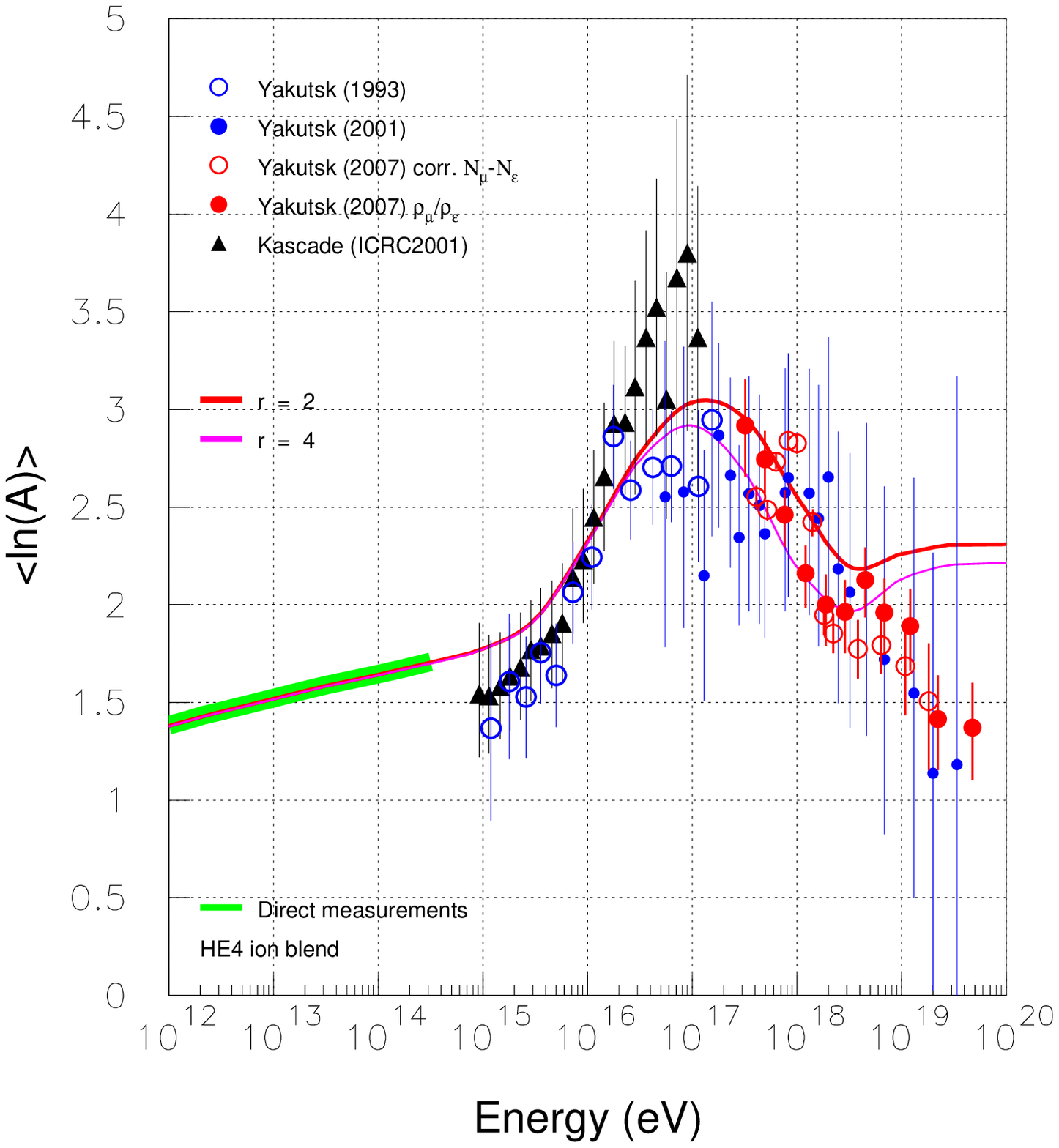}
\vspace{-1.6cm} \caption{ Comparison of the theoretical profile of
 $<$$ln(A)$$>$ (red curve) with the empirical one extracted from the data
 of the Yakutsk experiment (blue small circles).}
\label{fig:largenenough}
\vspace{-0.7cm}
\end{figure}

Figure 20  is an illustration of how ion abundances  of the
extradisc component are affected by small variations of the spectral
indices at the sources for cosmic ions migrating from the disc into
the extradisc space. The abundances of H, He and Fe ions in the
extradisc volume at $10^{19}$ eV are taken with the arbitrary ratios
of 1.5, 1 and 0.33 (see fig. 20). Note that the three black curves
(H, He and Fe) in fig. 20 are the same reported in figure 16 except
for a different normalization. The Helium index $\gamma_{He}$ is
taken as a reference value (any value, including $\gamma_{He}$=0)
while the Fe and H indices are varied by small amounts of 0.02,
0.04, etc. with respect to $\gamma_{He}$ , in order to map how ion
abundances vary with energy. The result is displayed in figure 20 by
the grid of red and blue curves  in the interval
$10^{12}$-$5\times10^{19}$ eV. For example,  with a change  of 0.08
in the index, the H/Fe abundance ratio of 19.9 (arbitrary value) at
$10^{12}$ eV becomes 0.23 at $10^{19}$ eV, a factor 86.5 minor. Such
a figure represents a large variation of the chemical composition.
Note that at very high energy, above $10^{19}$ eV, the abundance
ratios outside the disc volume are the same existing at the
cosmic-ion sources in the disc, since particle displacement is
unaffected by nuclear collisions,  due to the small grammage
traversed.

Measurements of the spectral indices in the energy region $10^{11}$
-$10^{15}$ eV acquire a fundamental importance because their values
determine the chemical composition at very high energy. The
postulate of $\it {Constant}$ $\it {Spectral}$ $\it {Indices}$,
incorporating the balloon and satellite data below $10^{15}$ eV,
relates the chemical composition at low energy to that at very high
energy by logical necessity avoiding any $\it {ad}$ $\it {hoc}$
mechanisms, at small, arbitrary  energy bands, to explain the
chemical composition of the cosmic radiation.


\begin{figure}[htb]
\vspace{-0.3cm}
\includegraphics [angle=0,width=8cm] {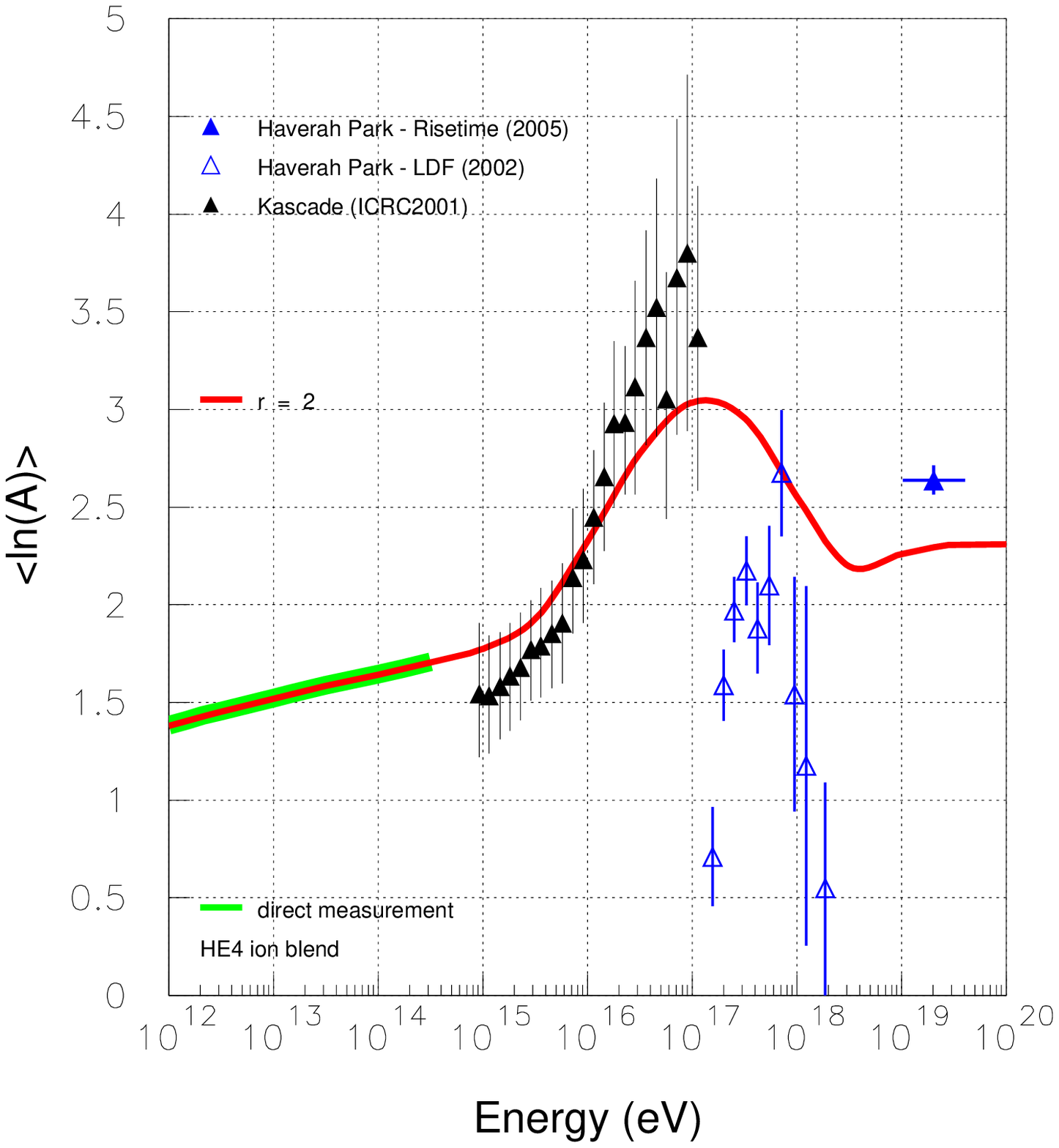}
\vspace{-1.6cm} \caption{ Comparison of the theoretical profile of
 $<$$ln(A)$$>$ (red curve) with the empirical one extracted from the data
 of the Haverah Park experiment (blue triangles).}
  \label{fig:largenenough}
\vspace{-0.7cm}
 \end{figure}


\section {The Chemical composition of the extradisc component }

\par The chemical composition of the extradisc component
is radically different from that of the disc component. Heavy ions,
escaping  from the disc volume and penetrating through the disc of
the Milky Way from its exterior,  pay twice a tribute to larger
nuclear interaction cross sections with respect to light ions.
Therefore,  the heavy-to-light  ion flux ratio inevitably augments
above $10^{15}$ eV up to the ankle energy region.

Figures 21 and 22 report the energy spectra of the extradisc ions
reaching the solar cavity for the two $LE$ and $HE4$ blends.
Firstly,  note that a simplified situation takes place above the
maximum of the Fe energy spectrum located at $2\times$$10^{19}$ eV.
The ion abundances above this energy coincide with those  at
$10^{12}$ eV reported in Table 1. Above
 $2\times$$10^{19}$ eV the grammage is so small  that
nuclear interactions in the
 interstellar medium do not alter
the chemical composition of the extradisc component existing at the
outskirts of the disc volume. The complete spectrum in figure 21 and
that in figure 22 are normalized at the arbitrary energy of
$10^{19}$ eV to the galactic flux of $5.35\times$$10^{14}$ ($LE$)
and $3.91\times$$10^{14}$ ($HE4$) particles/$m^2$ sr s $eV^{1.5}$.

Figure 23 shows ion fractions versus energy of the extradisc
component for 3 groups of ions: light, intermediate and heavy being,
respectively, (H+He), (CNO + Ne-S) and (Ca + Fe). This ion partition
is adopted in some experiments. Let us now analyze the ion fraction
profiles starting from the extreme high energy where the situation
is simpler.

Just above $2\times 10^{19}$ eV  ion abundances of the extradisc
component reaching the solar system are equal to those  of the
cosmic-ray component present in the parent galaxy,  because particle
displacement through the disc (escaping and entering) do not alter
ion abundances. Around $2\times10^{19}$ eV terminates the increase
of 2 orders of magnitude of the (Ca+Fe) ion fraction. The (Ca+Fe)
fraction above $5\times10^{18}$ eV does not surpass 17 per cent,
even with an increase of a factor 66 from the low level flat
fraction, below $5\times10^{16}$ eV. The intermediate ions, i.e. the
sum of CNO and (Ne-S) group of nuclei, has a similar pattern with a
step, between the low and high level,  of a factor 25, which
constitutes 31 per cent of the extradisc component above
$5\times10^{18}$ eV.

 From the profiles of the ion
fractions shown in figure 23 or from the more detailed ion spectra
shown in figures 21 and 22,  the $<$$ln(A)$$>$ of the extradisc
component is directly calculated; it is shown in figure 24 (pink
curve, $HE4$ blend). The same figure 24 shows, for visual
comparison, the corresponding $<$$ln(A)$$>$ resulting from the disk
component (blue curve, blend $HE4$), which has a quite different
pattern. The $<$$ln(A)$$>$ for the $LE$ blend of $I_{ed}$ has a
quite similar pattern (fig. 10). The two $<$$ln(A)$$>$ profiles of
the disc and the extradisc components in figure 24 are normalized at
the same cosmic-ray intensity at Earth.
\par The $<$$ln(A)$$>$ of the extradisc component exhibits
a constant value below $5\times10^{16}$ eV, followed by a rising
trend in the interval $5\times10^{16}$-$2\times10^{19}$ eV with an
approximate constant slope. Above $2\times10^{19}$ eV the
$<$$ln(A)$$>$ stabilizes to the constant value of 2.1. The physical
phenomena shaping this silhouette have already been discussed. A
simplified, instant, qualitative comprehension might be depicted  in
terms of grammage encountered by the cosmic rays during the
displacement in the Galaxy. When the grammage is large (see figure
16 ref. \cite{ginocchio}) heavy nuclei are destroyed, when the
grammage is minimum (0.006 g/$cm^2$) heavy nuclei travel undisturbed
like light nuclei and the persists unchanged regardless of the
energy. In the intermediate region $5\times10^{16}$-$2\times10^{19}$
nuclear cross sections and geometrical factors compete and
interplay, forging the chemical composition with the stable, rising
trend shown in figure 24. \quad

\par When the disc and extradisc components are combined together for different
values of $r$=$I_{ed}$/$I_{d}$,  it results the surprising
$<$$ln(A)$$>$ profiles shown in figure 25. For $r$=$1$ the flux in
the solar cavity is $1.8\times10^{14}$ particles/($m^2$ sr s
$eV^{1.5}$)  at $10^{19}$ eV for both disc and extradisc components.
The surprising aspect of the $<$$ln(A)$$>$ profiles in figure 25
resides in the fact that they thoroughly resemble to the profile of
the experimental data on $<$$ln(A)$$>$ extracted from the
measurements of $X_{max}$ of the Auger experiment.

Flux measurements of the cosmic radiation at Earth  imperatively
constraint $r$, the  maximum value of the extradisc component,  as
evident in figure 26.  The sum of the fluxes of the disc and
extradisc components is shown in figure 26 along with flux data of
two experiments, taken as an example. Magnifying the extradisc
component to match the Auger data on
 $<$$ln(A)$$>$ would entail a conflict with the observed fluxes.
\par  Notice that the theoretical spectra in figure 26 intrinsically
exhibit the adequate change in the slope around $4\times10^{18}$ eV,
from 3.2 to 2.7, which constitutes the ankle. Similarly,  around
$10^{15}$ eV, the slope of the complete spectrum changes from 2.7 to
about 3.0 which is a characteristic features of the knee.

\par With the assumptions adopted in this paper
any types of extradisc component generate a rather light chemical
composition below $10^{17}$ eV due to the dominant role of nuclear
cross sections suffered by heavy ions  while propagating from
periphery to disc core.


\begin{figure}[htb]
\vspace{-0.3cm}
\includegraphics [angle=0,width=8cm] {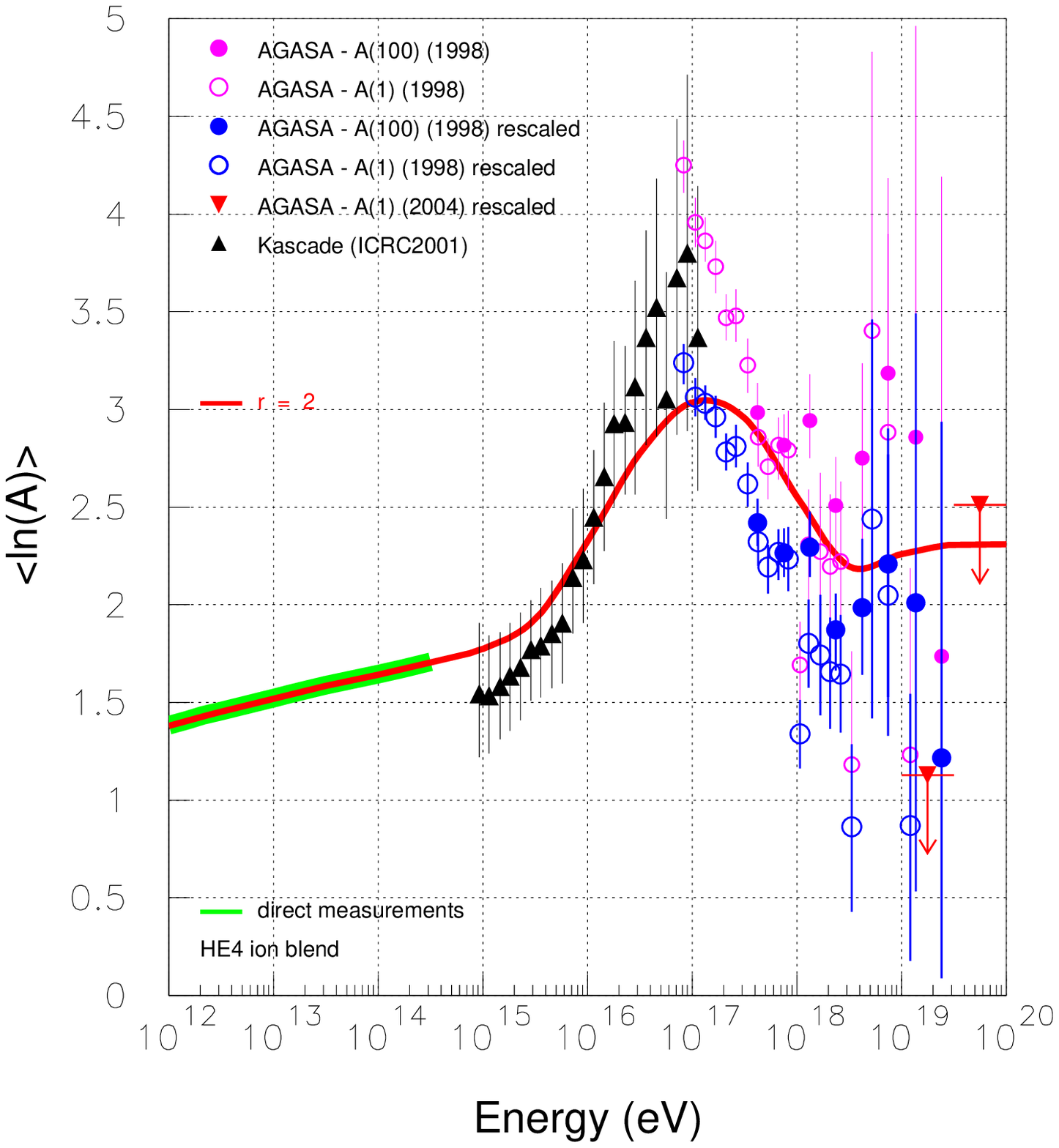}
\vspace{-1.6cm} \caption{ Comparison of the theoretical profile of
 $<$$ln(A)$$>$ (red curve) with the empirical one extracted from the data
 of Agasa experiment (blue and pink data points).}
  \label{fig:largenenough}
\vspace{-0.7cm}
 \end{figure}


\begin{figure}[htb]
\vspace{-0.3cm}
\includegraphics [angle=0,width=8cm] {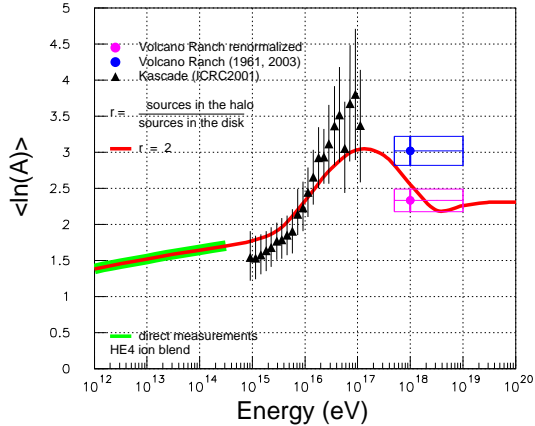}
\vspace{-1.6cm} \caption{ Comparison of the theoretical profile of
 $<$$ln(A)$$>$ (red curve) with the empirical one extracted from the data
 of Volcano Ranch experiment (blue cross in the rectangle). The data
 point
 corrected according to the $\it {Theory}$ $\it {of}$ $\it {Constant}$
  $\it {Spectral}$ $\it
{Indices}$ is represented by a pink cross in the rectangle.}
  \label{fig:largenenough}
\vspace{-0.7cm}
 \end{figure}


\begin{figure}[htb]
\vspace{-0.3cm}
\includegraphics [angle=0,width=8cm] {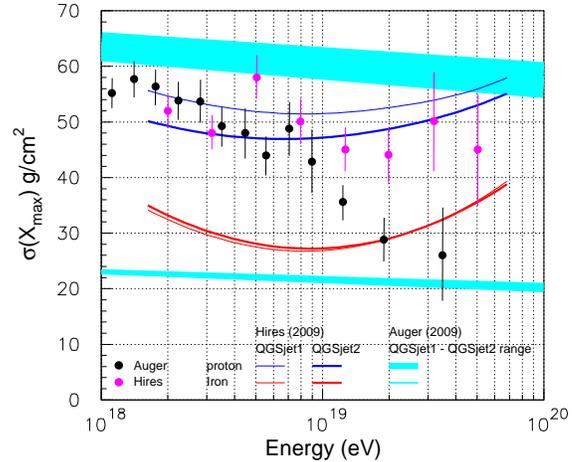}
\vspace{-1.6cm} \caption{Measurements of the chemical composition of
the cosmic radiation by HiRes (pink dots) and Auger (black dots)
experiments using the mean width of the longitudinal profile of the
fluorescence light generated in atmospheric showers. The theoretical
mean width in $g$/$cm^2$ versus energy evaluated by Heck \cite{heck}
for protons and Fe nuclei with  QGSjet and Sibyll algorithms  is
also shown as turquoise bands. } \label{fig:largenenough}
\vspace{-0.7cm}
\end{figure}

\begin{figure}[htb]
\vspace{-0.3cm}
\includegraphics [angle=0,width=8cm] {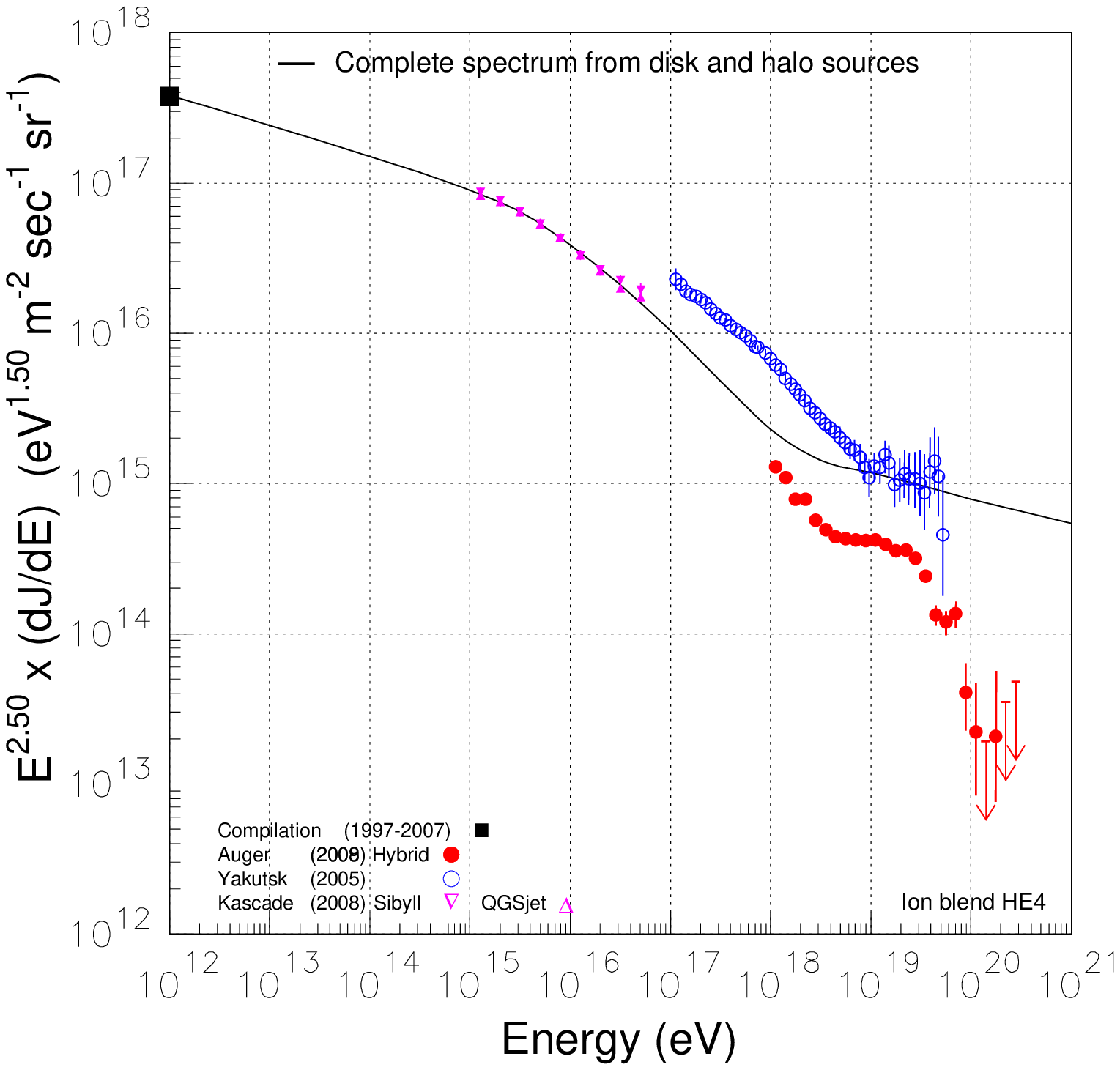}
\vspace{-1.6cm} \caption{Evidence for the present uncertainties in
cosmic-ray flux measurements in different experiments using giant
terrestrial cascades in air above $10^{17}$ eV. The spectrum (black
curve) resulting from the $\it {Theory}$ $\it {of}$ $\it {Constant}$
$\it {Spectral}$ $\it {Indices}$ is normalized  at $10^{12}$ eV
(black square) with a flux of $3.79 \times 10^{17}$ particles/($m^2$
sr s $eV^ {1.5}$) and it joins the Kaskade data in an excellent
accord.}
  \label{fig:largenenough}
\vspace{-0.7cm}
 \end{figure}

\section {Comparison between theoretical and measured $<$$ln(A)$$>$ with the
disc and extradisc components}

\par In the following the experimental data on $<$$ln(A)$$>$ are
compared with the profile derived from the theory in the condition
$r$=$2$.  Note that the $<$$ln(A)$$>$ in the  the condition $r$=$1$
or $r$=$3$ would describe the experimental data as well, due to the
large systematic uncertainties inherent  to the hadronic models
describing atmospheric showers.  The value $r$=$2$ is regarded here
as a adequate compromise between cosmic-ray intensity (see fig. 26)
and the depth of the minimum of the $<$$ln(A)$$>$ profile (see fig.
25).

\par In all the subsequent figures 27, 28, 29, 30, 31, 32,
and 33,  where high energy data on $<$$ln(A)$$>$ above  $10^{17}$
 eV are examined and compared with the theory,
 the $<$$ln(A)$$>$ determined by the Kascade experiment in the
range $10^{15}$-$10^{17}$ eV is displayed as adequate, reference
measurement.

 From the Kascade Collaboration
three different procedures to
 determine $<$$ln(A)$$>$  are available called here Sybill algorithm, QGSjet
algorithm and deconvolution method. Data of the first two methods
are in figure 3 and 27 while those of the deconvolution method,
which covers the largest range, i.e.
$9\times10^{14}$-$1.4\times10^{17}$ eV, are in figure 28. The
resulting  $<$$ln(A)$$>$  profiles from the three methods of
measurement by Kascade are simultaneously shown elsewhere (see
figure 24 of ref. \cite{nota2flussidipro}) exhibiting a coeherent
silouhette in the entire interval $10^{15}$-$10^{17}$ eV.

\par Recent measurements of $X_{max}$ of the Auger Collaboration
converted into $<$$ln(A)$$>$ by the reference profiles for H and Fe
shown in figure 14 are given in figure 27. Two predicted
$<$$ln(A)$$>$ profiles according to the $\it {Theory}$ $\it {of}$
$\it {Constant}$ $\it {Indices}$ with $r$=$2$ and $r$=$10$ are shown
in fig. 27.  In the condition $r$=$2$ the Auger data in the interval
$4\times 10^{17}$-$10^{19}$ eV are still below the theoretical
profile.

\par The $<$$ln(A)$$>$  extracted from the $X_{max}$ measured by the
Fly's Eye experiment \cite{occhidimosca} is shown in figure 28. The
increasing trend of  $<$$ln(A)$$>$  in the interval
$10^{17}$-$6\times10^{17}$ eV contrasts with the opposite trend of
HiRes and Agasa-Akeno experiments (see figs. 29 and 32); it also
disagrees with a flat silouhette observed by Yakutsk in the same
energy interval,  and with a value of 3.5 measured by Kascade. The
behavior of $<$$ln(A)$$>$ in figure 28 is reminescent of that
observed by the Haverah Park experiment (see fig. 31) in the same
energy interval, where a lack of uniformity in the detector
acceptance is suspected.

\par  The profile of $<$$ln(A)$$>$  extracted from the $X_{max}$
measured by the HiRes experiment  is shown in figure 29. The HiRes
data in figure 29 include HiRes Prototype \cite{hiresabuzay} , HiRes
Stereo \cite{hiresoriginal}   and a data revision which takes into
account acceptance correction and detector performance
\cite{belzsalina2008}. Around $4\times 10^{18}$ eV data reported in
figure 29 have a mean value of $<$$ln(A)$$>$ of about 1. This tiny
value would correspond to the minimum observed by Auger at
$(2-4)\times10^{18}$ eV as far as a vague increase of $<$$ln(A)$$>$
would delineate in the band $5\times 10^{18}$-$4\times 10^{19}$ eV.
The chemical composition extracted from the $X_{max}$ measured by
HiRes is lighter than that resulting from the highest permissible
values of the parameter $r$ (dominance of extradisc component) in
the interval $10^{17}$-$10^{19}$ eV. It is also lighter, by about
one unit of $<$$ln(A)$$>$,  than that of Auger, Fly s' Eye, Haverah
Park and Agasa experiments (see figure 4 of ref.
\cite{bottomaggiore}.)

\par The profile of $<$$ln(A)$$>$ extracted from the $X_{max}$ measured
by the Yakutsk Collaboration \cite{iachuscho} is shown in figure 30.
There is a global, remarkable agreement with the theory in the huge
interval $10^{15}$-$10^{19}$ eV characterized by a bell-shaped
profile with a maximum of $<$$ln(A)$$>$ of 3 around the energy of
$2\times 10^{17}$ eV. Above $5\times 10^{18}$ eV data would suggest
the lightning of the chemical composition. Notice, by contrast, that
the chemical composition of measured by  Auger in the interval
$5\times 10^{18}$-$4\times 10^{19}$ eV would indicate the opposite
trend: the
 $<$$ln(A)$$>$ becomes heavier according to the data shown in figure 27.

\section {Comparison between theoretical and measured $<$$ln(A)$$>$ in Haverah
Park,
 Agasa and Volcano Ranch}

\par The partition of the cosmic nuclei in two groups,
light and heavy, with fractions $F_p$ and $F_{Fe}$ respectively,
characterizes the chemical composition extracted from Haverah Park ,
Agasa and Volcano Ranch data samples. This bi-modal data analysis,
though appropriate for some detectors,  has a clear bias since it
gives intrinsically higher $<$$ln(A)$$>$ than  realistic ion
groupings as demonstrated below.

The chemical composition measured by the Haverah Park experiment is
reported in figure 31. Ruling out the  data points at the extreme
energies, below $2\times10^{17}$ eV and above $10^{18}$ eV, where
detector acceptance for atmospheric showers might have been
deteriorated (See Section 3 ref. \cite{haverahpark}), the agreement
with the theoretical $<$$ln(A)$$>$ profile in the condition $r$=$2$
is excellent. For protons and Fe nuclei the differences in the
theoretical $X_{max}$ profiles shown in figure 12 for Sibyll and
QGSjet codes are  small in the interval $10^{17}$-$10^{18}$ eV,  and
so the agreement between theory and Haverah Park data is quite
stable. It has been remarked \cite{watsonfpffe} the critical role of
hadronic models in the partition of all cosmic-ray nuclei in two
classes, light ($F_p$) and heavy ($F_{Fe}$).  Adopting the code
$QGSjet98$ it results a proton fraction $F_p$ of .34 while the code
$QGSjet01$ gives $F_{Fe}$ = $0.48$ \cite{watsonfpffe}.

 \par  Data samples collected by the
Agasa Collaboration have been revisited in 2003 \cite{dova-watson}
in order to determine the $X_{max}$ of the cosmic radiation in the
range $10^{17}$-$10^{20}$ eV. The chemical composition resulting
from this data re-elaboration  \cite{dova-watson} is shown in figure
32 by blue squares. The global profile of the empirical
$<$$ln(A)$$>$ in figure 32 would resemble to that measured by Auger
in a significant aspect: $<$$ln(A)$$>$ decreases above the energy of
$10^{17}$ eV up to $(2-3)\times 10^{18}$ eV, where a minimum is
attained. In the interval $2\times 10^{18}$-$2\times 10^{19}$ eV,
with some imagination, an increasing trend would faintly delineate,
being in accord with the $<$$ln(A)$$>$  profile of the Auger
experiment above $3\times10^{18}$ eV (see fig. 27).

\par  The partition of atmospheric showers in two groups according
to the quoted re-analysis of the Agasa data \cite{dova-watson} has
been here refined with 6 particles (or 6 ion groups) giving the
outcome shown in figure 32 by pink square data points. In this
refinement the $\it {Theory}$ $\it {of}$ $\it {Constant}$ $\it
{Spectral}$ $\it {Indices}$  with the $HE4$  blend  has been used.
The resulting
 $<$$ln(A)$$>$ profile (pink squares in fig. 32) is
 systematically shifted downward by about one unit of $<$$ln(A)$$>$.
 To appreciate numerically such a shift,
 just caused by the
refinement with 6 ion groups instead of 2, consider that a Fe
fraction of $0.60$ implies a $<$$ln(A)$$>$ of 2.41 ($A$ = $56$). The
$HE4$ blend at the energy of $10^{19}$ eV for the 6 ions H, He, CNO,
Ne-S, Ca and Fe with fractions, respectively, of .22, .21, .16, .16,
0.02 and 0.23 gives a $<$$ln(A)$$>$ of 2.26. The $LE$ blend at
$10^{19}$ eV with the ion fractions of .08, .41, .10, .09,  0.01 and
0.31 gives a $<$$ln(A)$$>$ of 2.43. \footnote{Unfortunately, an
heavy ion fraction $F_{Fe}$ higher than 1 results in figure 4 of
ref. \cite{dova-watson}, which signals a logical incoherence
whatsoever in the data elaboration materializing an unphysical
situation. The incoherence is not easy to be eliminated since
reverberates elsewhere \cite{simpsonagasatesi}. Probably, an
additional correction of a few $g$/$cm^2$ in the $X^{cal}_{max}$
derived from the Mocca-Sibyll code would eliminate the Fe fractions
higher than 1 in fig. 4 of the quoted reference \cite{dova-watson}
or in fig. 4 of ref. \cite{dawson}. Note that this additional
correction  probably is unnecessary since the oversimplified data
elaboration with only two ion groups (instead of 6 or more) entails
a bias in $<$$ln(A)$$>$ which is clearly demonstrated in figure 32
or in figure 33 with the Volcano Ranch data. }

A consistency verification  of the chemical compositions derived
from the Agasa-Akeno data  \cite{akenodatiagasa} and that extracted
by $X_{max}$ measured by Fly's Eye has been performed in 1998
\cite{dawson}. Both the radial muon density from the cascade core
$\rho_{\mu}$ and the $X^{cal}_{max}$ have been simulated with the
same hadronic code, Mocca-Sibyll. Calculations have been adapted to
Akeno (A1), Agasa (A100) and  Fly' s Eye detectors including muon
thresholds, trigger efficiencies and other instrumental constraints,
which are significant. Grouping all nuclei in two fractions, light
and heavy, so that $F_p$ + $F_{Fe}$=$1$, it results a fraction
$F_{Fe}$ greater than 1 for the Fly's Eye data around $10^{17}$ eV.
In order to remove this incoherence a shift of - 30 $g$/$cm^2$ in
the $X^{cal}_{max}$ is required. The interesting aspect of this
correction is that the theoretical profile $X^{cal}_{max}$ resulting
from Mocca-Sibyll code differs from that used in this paper (see
figure 12) by about 25 $g$/$cm^2$ at $10^{17}$ eV,  and similarly at
higher energies. Such a difference would alleviate, as a gratuity,
the gap between the $<$$ln(A)$$>$ predicted by the $\it {Theory}$
$\it {of}$ $\it {Constant}$ $\it {Spectral}$ $\it {Indices}$ with
r=2 and the Auger data (see fig. 27).

\par  An elaboration and revision of atmospheric shower data of Volcano
Ranch experiment operated in the period 1959-1964 has been performed
in 1975 using a bi-modal analysis \cite{dovamattiazziwa}.  Figure 33
shows the theoretical profile (red curve) along with the data point
of Volcano Ranch (blue rectangle) spanning the interval $5\times
10^{17}$-$10^{19}$ eV. The resulting $<$$ln(A)$$>$ has a mean value
of 3,  which indicates that the cosmic radiation at $10^{18}$ eV is
still dominated by intermediate and heavy nuclei. If the same
procedure of ion ponderation applied to the Agasa data is also
applied to Volcano Ranch data with the $HE4$ ion blend,  the mean
value of $<$$ln(A)$$>$ decreases from 3 to 2.4 (pink cross in the
rectangle) approaching the theoretical value of 2.5 as displayed in
figure 33.


\section{Reported measurements of the mean atmospheric depth
of the cosmic radiation by fluorescence light in HiRes}

Though the longitudinal profile of giant atmospheric showers
reconstructed by the fluorescence light is rather insensitive to the
hadronic models employed in the data analysis, it would seem that
some instrumental unknowns or uncontrolled measurement procedures
still plagues the determination of $X_{max}$ by fluorescence light.

Preliminarily, notice that experiments taking advantage of the
fluorescence light in the measurements of $X_{max}$  like HiRes do
not agree one another. The $X_{max}$ observed by the HiRes
experiment \cite{belzsalina2008} is larger than that measured by
Auger \cite{augerloga} which uses the fluorescence light as well. As
an example, the Fly' s Eye data on $<$$ln(A)$$>$ \cite{occhidimosca}
in the energy decade $3\times 10^{17}$-$3\times 10^{18}$ eV shown in
figure 28 differ from those of HiRes Prototype (figure 29) by one
unit of $<$$ln(A)$$>$ which is quite large by any standard,
theoretical, empirical or instrumental.

\par The measurements of  $X_{max}$ by  florescence light demands the
interpolation of the longitudinal profile by an appropriate function
$f_{LP}$ which has a characteristic width denoted here
$\sigma({X_{max}})$. The average value of the  distribution of
$\sigma({X_{max}})$ at a given energy is a measurement of the
chemical composition.  The observed $\sigma({X_{max}})$ may be
cross-check method in the measurements of the chemical composition
with the same data samples.

Presently (2009),  the available measurements of $\sigma({X_{max}})$
of the  HiRes Collaboration \cite{hiresnorvegia} disagree with the
analogous  data of the Auger experiment \cite{sigmaauger2009} as
shown in figure 34. Note that the theoretical $\sigma({X_{max}})$
profiles shown in figure 34 derive from the same sample of the
simulated cascades  used to calculate the $X_{max}$ profiles of
figure 12. The $\sigma({X_{max}})$ profile measured by HiRes is
rather flat around the value of 50 $g$/$cm^2$ in the interval
$10^{18}$- $5\times10^{19}$ eV while Auger data (full black dots)
exhibit a clear decreasing profile from a maximum of 58 $g$/$cm^2$
down to a minimum of 22 $g$/$cm^2$. Imagine a linear scale of energy
between $10^{18}$- $5\times10^{19}$ eV in figure 34 and the
disagreement between the two experiments dilates,  since the initial
agreement below $4\times10^{18}$ eV is confined in less than 5 per
cent of the explored energy band.

At the energy of $10^{17}$ eV the profile of $<$$ln(A)$$>$ measured
by Kascade joins that of the HiRes experiment \cite{hiresabuzay}
(fig. 29). The average value of the $<$$ln(A)$$>$ of the Kascade
experiment is about 3.1 (average value of 3 methods of measurements)
while that of HiRes is close to 2.5. It is important to emphasize
the disagreement, or the potential disagreement,  between these two
experiments on $<$$ln(A)$$>$ around $10^{17}$ eV. The knees of the
light ions have been observed by a number of experiments above
$10^{15}$ eV \cite{{macroeastop},{kascadekampert}}. This fact
necessarily implies an increase of $<$$ln(A)$$>$ above $10^{15}$ eV,
since the disappearance of light ions automatically enhances the
heavy ion fraction. Knowing that at $10^{15}$ eV  the value of
$<$$ln(A)$$>$ is certainly comprised between 1.74 and 1.78 as
inferred by extrapolating balloon and satellite data
\cite{orandellogdea}, any estimate of $<$$ln(A)$$>$ at $10^{17}$ eV
leads to the value of 3.1 and not a lower value as HiRes Prototype
data (2004) \cite{hiresabuzay} around $10^{17}$ eV in figure 29
would suggest.

As apparent in figure 29 the mean values of $<$$ln(A)$$>$
 of the HiRes experiment after the acceptance correction \cite{belzsalina2008}
 are globally displaced downward by some  0.5 units.
 The shift is surprisingly large compared to the reported error bars before
 and after acceptance corrections \cite{belzsalina2008}. The large
 excursions in $<$$ln(A)$$>$ (blue dots) also signal that the global detector performance
 suffered from residual uncontrolled unknowns.

The correction of the detector acceptance  \cite{belzsalina2008}, a
 routine effect
 applied after 3 years from the original measurement \cite{hiresoriginal}
 does not preclude future data re-analysis.

In synthesis, the HiRes data pertaining the chemical composition of
the cosmic radiation disagree \footnote{ The resonance of this
conclusion tunes up with the criticism expressed by A. A. Watson
\cite{velenowatson} on Fly's Eye and HiRes experiments: $\ll$.... By
contrast, with florescence detectors the aperture continues to grow
with energy and remains considerable uncertainty about the HiRes
aperture. Extensive Monte Carlo calculations must be made to
establish it and these make assumptions about the slope of the
spectrum and about the primary mass, with protons being assumed on
the basis of the claims from the Fly' Eye and HiRes experiments
ignoring often contradictory evidence from other experiments without
discussion $\gg$.}: (1) with the Auger data on $X_{max}$; (2) with
the Auger data on $\sigma({X_{max}})$ (see figure 34 and the related
text); (3) with those of all other experiments on $<$$ln(A)$$>$
above $10^{17}$ eV (Volcano Ranch, Haverah Park, Akeno, Agasa, Fly'
s Eye and Yakutsk); (4) with the Kascade data at $10^{17}$ eV.
Theoretically, a bias exists in the conversion of $X_{max}$ into
$<$$ln(A)$$>$ made by the HiRes Collaboration because, out of all
current versions of hadronic codes adopted to simulate nuclear
interactions in air (e.g. QGSjet, Sibyll, etc.), those adopted by
HiRes tend to generate protons and depress heavy ions (see fig. 8 of
ref. 10 and fig. 4 ref. \cite{bottomaggiore}).


\begin{figure}[htb]
\vspace{-0.3cm}
\includegraphics [angle=0,width=8cm] {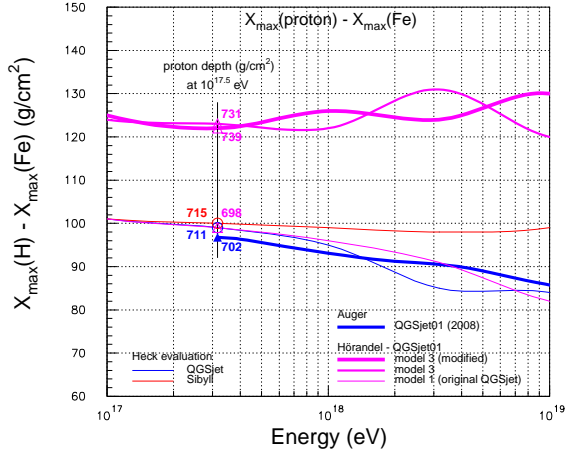}
\vspace{-1.6cm} \caption{Theoretical difference of the atmospheric
depth for protons and Fe nuclei according to a number of
evaluations. The model QGSjet-03 (thick pink curve)
\cite{orandellogdea} would reduce the discrepancy between Auger data
on $<$$ln(A)$$>$ and the corresponding profile derived from the $\it
{Theory}$ $\it {of}$ $\it {Constant}$ $\it {Indices}$.}
\label{fig:largenenough} \vspace{-0.7cm}
\end{figure}


\section{Conclusions.}

 The comparison of the
 chemical composition of the cosmic radiation expressed by $<$$ln(A)$$>$
derived from the $\it {Theory \quad of \quad Constant \quad Indices
}$ with the observed $<$$ln(A)$$>$  extracted from the Auger data on
 $X_{max}$ dictates the necessity of introducing an extradisc
 component of
the cosmic radiation above $10^{17}$ eV. In Sections 1 and 2 it is
explained why this component is better termed $\it {extradisc}$ $\it
{component}$,  $I_{ed}$. The variable $r$ is a free parameter of the
theory defined as the extradisc-to-disc flux ratio in the solar
cavity at $10^{19}$ eV, e.g. $r$=$I_{ed}$/$I_{d}$.

  From the ensemble of the profiles  of $<$$ln(A)$$>$ extracted
from the measurements of
  $X_{max}$ of the Auger, Yakutsk, Fly' s Eye, Agasa, Akeno, Haverah
Park and Volcano Range experiments it emerges a global accord with
the theory.

In the energy interval $10^{15}$-$10^{17}$ eV there is an excellent
agreement with the observations of Eas-top, Kascade, Tunka (see fig.
3) and other experiments.
  Below
$10^{17}$ eV  the extradisc component is a negligible fraction of
the disc component $I_d$, and as a consequence, the theoretical and
empirical analysis reported in the companion paper [8] remains
valid.

Because of the unknowns entailed in the conversion of $X_{max}$ into
$<$$ln(A)$$>$ above $10^{17}$ eV,   the accord between the
theoretical profile and data in figures 27, 28, 30, 31, 32 and 33 is
more than satisfactory. Values of $r$= $I_{ed}$/$I_{d}$ higher than
2 would alleviate the gap between the empirical profile of
$<$$ln(A)$$>$ and the theory, as shown in figure 27. Presently,
values of $r$ higher than 2 are neither convincing nor necessary,
due to large systematic errors in the experimental data. The last
two figures 35 and 36 intend to illustrate this position.

\par Figure 35 depicts in a flash the present experimental
uncertainties in flux measurements quoting arbitrarily data from
Kascade, Yakutsk and Auger experiments. Fluxes from other
experiments do not agree as well. Such uncertainties impede to
identify  a more precise and reliable value of the ratio $r$ in
figure 26 on an empirical basis,  being $r$=$2$ the present adequate
global estimate.

Let be $X^{cal}_{max}$ (H) the theoretical atmospheric depth for
protons, $X^{cal}_{max}$ (Fe) that for Fe nuclei and $D_{max}$ the
difference between the two depths i.e. $D_{max}$ = $X^{cal}_{max}$
(H)- $X^{cal}_{max}$ (Fe) at a given energy $E$. Error sources
generating differences in  $<$$ln(A)$$>$ in theories and in
experiments might arise both from $X^{cal}_{max}$ (H) and  $D_{max}$
as well. Ideally, different hadronic models may result in a
combination of $X^{cal}_{max}$ (H) and $D_{max}$ having the same
$<$$ln(A)$$>$. Figure 36 reports the $D_{max}$   versus energy in
some current hadronic models and,
 at the arbitrary energy of $10^{17.5}$ eV, the $X^{cal}_{max}$ (H).
  Figures 36 and 12 show that the
$X^{cal}_{max}$ (H)  and $D_{max}$ profiles used in this paper for
the conversion of  $X_{max}$ into  $<$$ln(A)$$>$ are quite similar
to those adopted by others (for example, Auger).
\par Notice that the
hadronic code called QGSjet model-03 \cite{orandellogdea}, having a
larger $D_{max}$ of 120-130 $g$/$cm^2$ in the relevant range
$10^{17}$-$5\times 10^{19}$ eV and a deeper $X^{cal}_{max}$ (H) of
739 $g$/$cm^2$  at $10^{17.5}$ eV (see figure 36), it would shift
upward the simulated $X_{max}$ profiles in figure 12 by 9 $g$/$cm^2$
at $10^{17.5}$ eV. The same correction applied to the Auger data in
figure 3 (or figure 14) would shift upward by 0.57 units of
$<$$ln(A)$$>$ (9 $g$/$cm^2$ for $r$=$0$ or 11 $g$/$cm^2$ for
$r$=$2$) at $4.12\times10^{17}$ eV alleviating the gap between Auger
data and theory. As far as uncertainties in hadronic models above
$10^{17}$ eV used in the analysis of giant atmospheric cascades
remain large, the determination of a  value of $r$ higher than 2
based on the $<$$ln(A)$$>$ data would appear premature.  This same
conclusion is suggested by the flux measurements in figure 35.

Let us finally mention the systematic disagreement between the
$<$$ln(A)$$>$ extracted from the $X_{max}$ measured by the HiRes
experiment and  theory,  amounting to about one unit in the entire
interval $10^{17}$-$5\times 10^{19}$ eV (see fig. 29). The
examination of the HiRes data on $<$$ln(A)$$>$ made in Section 9
suggests a number of inconsistencies with the data of numerous
experiments and with different versions of the HiRes experiment
itself (HiRes Prototype, HiRes Stereo, data revision
\cite{belzsalina2008}).

\end{document}